\documentstyle[myemulateapj]{article}

\begin{document}
\submitted{Submitted March 18, 1999; to appear in \emph{The Astrophysical Journal Supplement Series}}
\title{The Type I\lowercase{a} Supernova 1998\lowercase{bu} in M96 and the 
Hubble Constant}

\author{Saurabh Jha, Peter M. Garnavich, Robert P. Kirshner, Peter
Challis, Alicia M. Soderberg, Lucas M. Macri, John P. Huchra, Pauline
Barmby, Elizabeth J. Barton, Perry Berlind, Warren R. Brown, Nelson
Caldwell, Michael L. Calkins, Sheila J. Kannappan, Daniel M. Koranyi,
Michael A. Pahre\altaffilmark{1}, Kenneth J. Rines, Krzysztof
Z. Stanek, Robert P. Stefanik, Andrew H. Szentgyorgyi, Petri
V\"ais\"anen, Zhong Wang, and Joseph M. Zajac}
\affil{Harvard-Smithsonian Center for Astrophysics, 60 Garden Street,
Cambridge, MA 02138}

\author{Adam G. Riess, Alexei V. Filippenko, Weidong Li, Maryam
Modjaz, \\ and Richard R. Treffers} \affil{Department of Astronomy,
University of California, Berkeley, CA 94720-3411}

\author{Carl W. Hergenrother}
\affil{Lunar and Planetary Laboratory, University of Arizona, Tucson,
AZ 85721}

\author{Eva K. Grebel\altaffilmark{1,2}}
\affil{Department of Astronomy, University of Washington, Seattle, WA 98195}

\author{Patrick Seitzer}
\affil{Department of Astronomy, University of Michigan, Ann Arbor, MI 48109}

\author{George H. Jacoby}
\affil{Kitt Peak National Observatory, National Optical Astronomy
Observatories, Tucson, AZ 85726}

\author{Priscilla J. Benson}
\affil{Whitin Observatory, Wellesley College, Wellesley, MA 02181}

\author{Akbar Rizvi and Laurence A. Marschall}
\affil{Department of Physics, Gettysburg College, Gettysburg, PA
17325}

\author{Jeffrey D. Goldader}
\affil{Department of Physics and Astronomy, University of
Pennsylvania, Philadelphia, PA 19104}

\author{Matthew Beasley}
\affil{Department of Astrophysical and Planetary Sciences, University
of Colorado, Boulder, CO 80309}

\author{William D. Vacca}
\affil{Institute for Astronomy, University of Hawaii, Honolulu, HI 96822}

\author{Bruno Leibundgut and Jason Spyromilio}
\affil{European Southern Observatory, Karl-Schwarzschild-Strasse 2,
Garching, Germany}

\author{Brian P. Schmidt and Peter R. Wood}
\affil{Research School of Astronomy and Astrophysics, Australian
National University, Canberra ACT, Australia}

\altaffiltext{1}{Hubble Fellow}
\altaffiltext{2}{Also UCO/Lick Observatory, University of California,
Santa Cruz, CA 95064}

\begin{abstract}
We present optical and near-infrared photometry and spectroscopy of
the type Ia SN 1998bu in the Leo I Group galaxy M96 (NGC 3368). The
data set consists of 356 photometric measurements and 29 spectra of SN
1998bu between UT 1998 May 11 and July 15. The well-sampled light
curve indicates the supernova reached maximum light in $B$ on UT 1998
May 19.3 (JD $2450952.8 \pm 0.8$) with $B = 12.22 \, \pm \, 0.03$ and
$V = 11.88 \, \pm \, 0.02$. Application of a revised version of the
Multicolor Light Curve Shape (MLCS) method yields an extinction toward
the supernova of $A_V = 0.94 \pm 0.15$ mag, and indicates the
supernova was of average luminosity compared to other normal type Ia
supernovae. Using the \emph{HST} Cepheid distance modulus to M96
\markcite{tan95}(Tanvir et al. 1995) and the MLCS fit parameters for
the supernova, we derive an extinction-corrected absolute magnitude
for SN 1998bu at maximum, $M_V = -19.42 \, \pm \, 0.22$. Our
independent results for this supernova are consistent with those of
\markcite{sun99} Suntzeff et al. (1999). Combining SN 1998bu with
three other well-observed local calibrators and 42 supernovae in the
Hubble flow yields a Hubble constant, $H_0 = 64^{+8}_{-6} \; \rm{km \;
s^{-1} \; Mpc^{-1}}$, where the error estimate incorporates possible
sources of systematic uncertainty including the calibration of the
Cepheid period-luminosity relation, the metallicity dependence of the
Cepheid distance scale, and the distance to the LMC.
\end{abstract}


\section{Introduction}

Type Ia supernovae (SNe Ia) have recently assumed an elite status at
the top rung of the cosmic distance ladder. Comprehensive studies of
SNe Ia have established their general spectroscopic and photometric
homogeneity, along with quantifiable diversity. SNe Ia make reasonably
good standard candles
\markcite{kow68}\markcite{san93}\markcite{bra93}(Kowal 1968; Sandage
\& Tammann 1993; Branch \& Miller 1993) and excellent calibrated
candles
\markcite{phi93}\markcite{ham95}\markcite{ham96a}\markcite{rie95a}\markcite{rie96a}\markcite{tri98}\markcite{phi99}(Phillips
1993; Hamuy et al. 1995, 1996b; Riess, Press, \& Kirshner 1995a,
1996a; Tripp 1998; Phillips et al. 1999) which, combined with their
high intrinsic luminosity, makes them superb indicators of very large
distances and a powerful tool for cosmology \markcite{bra98}(Branch
1998, and references therein).

The downside is that SNe Ia are rare and fleeting, making them more
challenging to study than other astronomical distance
indicators. Nevertheless, recent applications of these ``standard
bombs'' have been numerous. Within a few hundred Mpc, they have been
used to measure the expansion of the Universe: with recession
velocities of their host galaxies and distances to individual SNe Ia,
the Hubble law has been tested to redshifts $z \simeq 0.1$ at high
precision \markcite{ham96a}\markcite{rie96a}\markcite{tam98}(Hamuy et
al. 1996b; Riess, Press, \& Kirshner 1996a; Tammann 1998). SNe Ia have
also been used to measure peculiar motions of galaxies and large scale
flows
\markcite{tam90}\markcite{mil92}\markcite{jer93}\markcite{rie95b}
\markcite{rie97a}\markcite{zeh98}\markcite{tam98}(Tammann \&
Leibundgut 1990; Miller \& Branch 1992; Jerjen \& Tammann 1993; Riess,
Press, \& Kirshner 1995b; Riess et al. 1997a; Zehavi et al. 1998;
Tammann 1998) as well as to provide constraints on the properties of
extragalactic dust (Riess, Press, \& Kirshner 1996b). At larger
distances, SNe Ia serve as standard clocks: cosmological time dilation
has been demonstrated by comparing light curves of distant supernovae
with light curves from those nearby
\markcite{lei96}\markcite{gol97}\markcite{rie97b}(Leibundgut et
al. 1996; Goldhaber et al. 1997), as well as from the relative rates
of spectral evolution (Riess et al. 1997b). Most recently, SNe Ia have
been used to measure luminosity distances at high redshift ($0.3
\lesssim z \lesssim 1.0$), from which the geometry and expansion
history of the Universe can be determined
\markcite{nor89}\markcite{per95}\markcite{sch98}(N$\o$rgaard-Nielsen
et al. 1989; Perlmutter et al. 1995; Schmidt et al. 1998). These
measurements have likely tolled the death knell of standard ($\Omega_M
= 1$) cold dark matter (CDM) cosmology, strongly disfavoring the
possibility of enough gravitating matter to flatten the Universe
\markcite{per98}\markcite{gar98a}(Perlmutter et al. 1998; Garnavich et
al. 1998a). Even more surprisingly, preliminary indications show that
the expansion is accelerating at the current epoch
\markcite{rie98a}\markcite{per99}(Riess et al. 1998a; Perlmutter et
al. 1999), inconsistent with the idea that ordinary matter is the
dominant component of the Universe's energy density. Alternatives such
as a cosmological constant, or other forms of energy density similar
to it, seem to be favored \markcite{gar98b}\markcite{per99}(Garnavich
et al. 1998b; Perlmutter et al. 1999).  Combined with constraints from
the cosmic microwave background anisotropy power spectrum, the
supernovae have also given preliminary indications for a Universe with
zero global curvature
\markcite{gar98b}\markcite{whi98}\markcite{lin98}\markcite{teg98}
\markcite{efs99}\markcite{ptw99}\markcite{roo99}(Garnavich et
al. 1998b; White 1998; Lineweaver 1998; Tegmark 1999; Efstathiou et
al. 1999; Perlmutter, Turner, \& White 1999; Roos \& Harun-or-Rashid
1999).

These studies rely on SNe Ia only as bright, precise, \emph{relative}
distance indicators. However, the determination of the expansion rate
of the Universe, the Hubble constant, and the age of the Universe
require an accurate \emph{absolute} distance scale. Measuring absolute
distances to SNe Ia requires calibrating them through objects on lower
rungs of the distance ladder. The best distance indicator for this
remains what it has been since the days of Hubble himself: $\delta$
Cephei variable stars. Cepheids are bright enough to be studied in
nearby galaxies, including those which have hosted SNe Ia. In addition
they can be found in significant numbers, which allows for a precise
mean period-luminosity (PL) relation to be constructed and compared to
nearby samples. In general Cepheids in external galaxies are compared
to Cepheids in the Large Magellanic Cloud (LMC), whose distance is
calibrated through a variety of means (not always with the same
result, cf.  Section \ref{sectlmcdist}). The importance of Cepheids to
calibrate secondary distance indicators (including SNe Ia) prompted
the creation of the \emph{Hubble Space Telescope} (\emph{HST}) Key
Project on the Extragalactic Distance Scale, which has measured
Cepheid distances to a number of nearby galaxies
\markcite{fre94}\markcite{fre98}(Freedman et al. 1994, 1998 and
references therein). Another \emph{HST} effort has been underway to
measure Cepheid distances to galaxies that have hosted SNe Ia, to
calibrate them specifically \markcite{san92}\markcite{sah97}(Sandage
et al. 1992; Saha et al. 1997, and references therein). At present the
use of SNe Ia to measure the Hubble constant is limited by the paucity
of local calibrators, not by the number or precision of distances to
objects in the Hubble flow.

Nature has been kind by providing us with SN 1998bu in the Leo I Group
galaxy M96 (NGC 3368), for which a Cepheid distance had already been
obtained with \emph{HST} by a third group, \markcite{tan95}Tanvir et
al. (1995). In this paper we calibrate the absolute magnitude of SN
1998bu through extensive optical and near-infrared photometry and
spectroscopy. In \S 2 we describe our observations and reduction
procedure. In \S 3 we derive parameters of SN 1998bu, including the
extinction along the line of sight as well as a quantitative estimate
of the intrinsic luminosity of SN 1998bu compared to other SNe Ia. In
\S 4 we combine the properties of SN 1998bu with other
Cepheid-calibrated SNe Ia and SNe Ia in the Hubble flow to determine
the Hubble constant ($H_0$) and its statistical uncertainty. We
discuss our results, sources of systematic error, and implications for
the age of the Universe, $t_0$, in \S 5. Finally, we conclude and
summarize in \S 6. Independent observations and analysis of SN 1998bu
have been performed by \markcite{sun99}Suntzeff et al. (1999), and we
compare our analysis and results with theirs throughout the
paper. Infrared and optical spectra and uncalibrated light curves of
SN 1998bu have also been presented by \markcite{mei99}Meikle \&
Hernandez (1999).

\section{Observations and Analysis}

\subsection{Discovery}

SN 1998bu was discovered by the amateur astronomer M. Villi on UT 1998
May 9.9 on CCD images of M96 \markcite{vil98}(Villi 1998). The
supernova was located at $\alpha = 10^{\rm h}46^{\rm m}46\fs03$,
$\delta = +11\arcdeg50\arcmin07\farcs1$ (equinox 2000.0), about one
arcminute north of the host galaxy nucleus \markcite{nak98}(Nakano \&
Aoki 1998). At the Center for Astrophysics we monitor new, bright
supernovae spectroscopically with the Center for Astrophysics
F. L. Whipple Observatory (FLWO) 1.5-m Tillinghast reflector and FAST
spectrograph \markcite{fab98}(Fabricant et al. 1998), and
photometrically with the FLWO 1.2-m reflector in optical and
near-infrared passbands.  Our photometric observations of SN 1998bu
with the 1.2-m began on May 11.1, with the new STELIRCAM near-infrared
InSb array detector.  The discovery of SN 1998bu ocurred during lunar
bright time, and so the 1.5-m telescope was not equipped with the FAST
instrument until May 15.

High resolution spectra taken by \markcite{mun98}Munari et al. (1998)
with the Asiago Observatory 1.8-m telescope showed interstellar
\ion{Na}{1} D absorption from our Galaxy, as well as from M96 at a
heliocentric radial velocity of $744.8 \pm 0.3 \; {\rm km \;
s^{-1}}$. Low-disperson spectra of SN 1998bu were taken by
\markcite{mei98}Meikle et al. (1998) on May 12.9 and
\markcite{aya98}Ayani, Nakatani \& Yamaoka (1998) on May 14.5, which
revealed the supernova to be of type Ia about a week before maximum
light. Our first spectroscopic observations of SN 1998bu with the FLWO
1.5-m were taken on May 16.1.

A type Ia supernova in a galaxy whose Cepheid distance had already
been measured by \emph{HST} \markcite{tan95}(Tanvir et al. 1995)
provided a unique opportunity; it was the opposite of the usual case,
in which an \emph{HST} Cepheid distance to a galaxy is measured
specifically because the galaxy was an SN Ia host. Well-measured light
curves are the key to SN Ia distances, so we undertook extensive
photometric observations in the $UBVRIJHK$ passbands.

\subsection{Optical Photometry}

The plurality of our optical photometric observations (29 nights) was
obtained with the FLWO 1.2-m telescope + ``4Shooter'' CCD mosaic
camera \markcite{sze99}(Szentgyorgyi et al. 1999). The 4Shooter
consists of a 2x2 array of thinned, back-side illuminated,
anti-reflective coated Loral $2048^2$ CCD detectors, situated at the
f/8 Cassegrain focus. The pixel size is 15 \micron, yielding a scale
of $0\farcs335$ per pixel at the focal plane and a field of view of
approximately $11\farcm4$ on a side for each chip, with total sky
coverage of 0.15 deg$^2$. Our observations were taken in a 2x2 binned
mode, so that the resulting images were sampled at $0\farcs67$ per
pixel, well matched to the typical seeing (1\farcs5-$2\arcsec$ FWHM)
achieved at this telescope. All observations of SN 1998bu with the
mosaic were made on the same CCD (chip 1), which has the best
combination of cosmetic characteristics and quantum efficiency. Two
nights of observations were obtained on this telescope with the
``AndyCam'' instrument, a CCD camera with a single CCD, very similar
to those that make up the 4Shooter. Both instruments have good
ultraviolet and near-infrared response, which enabled us to make
observations in the Johnson $UBV$ and Kron-Cousins $RI$ bandpasses.
Our optical filters are constructed from Schott glass components, as
recommended by \markcite{bes90}Bessell (1990) for coated CCDs. The
FLWO $BVRI$ filter prescriptions are described by
\markcite{rie99}Riess et al. (1999); in general they match well the
prescriptions of \markcite{bes90}Bessell (1990), though the FLWO $I$
filter extends to somewhat longer wavelengths. The FLWO $U$ filter
transmission is also a good match to the \markcite{bes90}Bessell
(1990) UX specification.

Our FLWO 1.2-m observations of SN 1998bu are part of an ongoing
supernova monitoring program at CfA. Supernova observations are not
well suited to the scheduled time allocation procedure that is typical
at most telescopes. To follow supernovae, the time allocation
committee authorizes us to enlist the generous aid of the scheduled
observers, asking them to devote a small fraction of observing time
(usually limited to $\sim 20$ minutes per night) to the SN program. We
complement the monitoring observations with scheduled nights (usually
one night per month) to measure fainter objects and perform
photometric calibrations. We have been quite successful observing in
this mode; a set of 22 SNe Ia light curves garnered as a result of
this program has been presented by \markcite{rie99}Riess et
al. (1999).

The FLWO 1.2-m is equipped with an infrared instrument during bright
time, which provides useful IR supernova data (obtained in a similar
observing mode), at the price of bright-time gaps in our optical light
curves. For this object we made a special effort to minimize these
gaps by inviting observers at other institutions to participate. Six
epochs of optical photometric observations in $UBVRI$ were taken at
the Michigan-Dartmouth-MIT (MDM) Observatory 2.4-m Hiltner telescope
and direct imager, with the thinned, back-illuminated, 1024$^2$
``Charlotte'' CCD detector, located at the f/7.5 Cassegrain focus and
providing a $4\farcm7 \times 4\farcm7$ field-of-view at $0\farcs28$
per pixel. Further observations on two nights were obtained at the
Kitt Peak National Observatory (KPNO) 0.9-m telescope, with the T2KA
(2048$^2$) CCD detector at the f/7.5 Cassegrain focus, yielding a
$23\arcmin \times 23\arcmin$ field-of-view at $0\farcs68$ per
pixel. Target of opportunity observations were also carried out on
five nights during NOAO time at the WIYN Observatory 3.5-m telescope
with the S2KB (2048$^2$) CCD Imager at f/6.3 with $0\farcs20$ per
pixel and a $6\farcm8 \times 6\farcm8$ field-of-view.

Other sites also observed SN 1998bu and we report those data as well.
We include CCD data from the Whitin Observatory 0.6-m telescope at
Wellesley College (eight nights using a 1024$^2$ CCD at f/13.5, with a
scale of $0\farcs91$ per pixel), the Gettysburg College Observatory
0.4-m (twelve nights using a 1024$^2$ CCD at f/11, $0\farcs84$ per
pixel) and the 0.76-m Katzman Automatic Imaging Telescope (KAIT) at
Lick Observatory run by the University of California, Berkeley
(fifteen nights with a 512$^2$ CCD at f/8.2, $0\farcs63$ per
pixel). The detectors used in these observations were not as
blue-sensitive as the others, so only $BVRI$ images were taken.
Our observations of SN 1998bu continued until July 2 when it was too
close to the setting sun to provide good photometric data. In total
our optical photometric data set consists of 327 measurements of SN
1998bu.

\begin{figure*}
\plotone{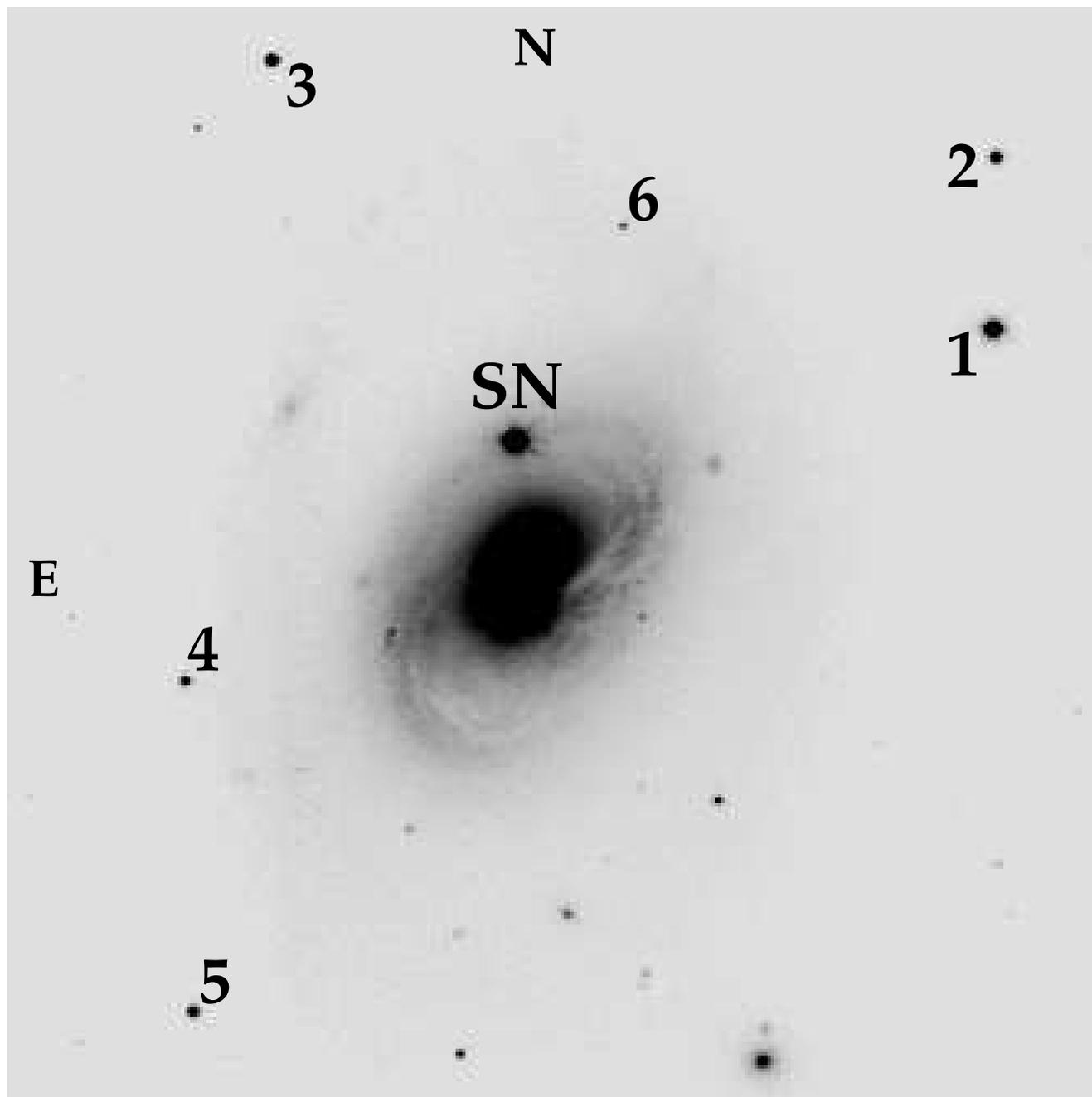}
\caption{SN 1998bu in M96 with local comparison
stars. The field is $7\farcm6 \times 7\farcm6$; north is up and east
is to the left. The image is a composite of several $V$-band
observations of SN 1998bu taken with the F. L. Whipple Observatory
1.2-m telescope in May of 1998. \label{figfield}}
\end{figure*}

All CCD observations were reduced (uniformly, beginning with the raw
data) in the standard fashion, with bias subtraction, dark current
subtraction (not necessary in most cases) and flat-field correction
using the IRAF CCDPROC package. Most of the observations were taken in
non-photometric conditions, so we have performed differential
photometry with a sequence of six comparison stars in the supernova
field, shown in Fig. \ref{figfield}. Comparison stars 1 and 2 were
calibrated on four photometric nights (two from the FLWO 1.2-m, one
each from the MDM 2.4-m and Kitt Peak 0.9-m). Stars 3, 4 and 5 were
not in the MDM field of view, so these were calibrated from three
nights. To calibrate the comparison stars into a local standard star
sequence, Landolt \markcite{lan92}(1992) standard fields providing
stars in a wide range of color were observed in $UBVRI$ over a wide
range in airmass. The supernova field was also observed in these
filters at an airmass within the airmass range of the standard star
observations.  These data were reduced and stellar instrumental
magnitudes were determined from aperture photometry using the APPHOT
package in IRAF. We then derived zero points and transformation
coefficients linear in airmass and color from the standard stars using
the prescription of Harris, Fitzgerald, \& Reed
\markcite{har81}(1981). This transformation was then applied to the
comparison star instrumental magnitudes to determine their standard
magnitudes. We treated observations from each photometric night
independently, and averaged the final standard magnitudes. Table
\ref{tabcstar} displays these mean standard magnitudes, along with the
error in the mean determined from the scatter of repeat
observations. We also list the comparison star identification numbers
from \markcite{sun99}Suntzeff et al. (1999). Our independent
photometry of the comparison stars agrees well with their results.

\begin{deluxetable}{ccccccccc}
\scriptsize
\tablecaption{Local standard star $UBVRIJHK$ magnitudes. \label{tabcstar}}
\tablehead{\colhead{Star} & \colhead{$U$} & \colhead{$B$} & 
\colhead{$V$} & \colhead{$R$} & \colhead{$I$} & \colhead{$J$} &
\colhead{$H$} & \colhead{$K$}}
\startdata
1 (S6) & 13.521(0.027) & 13.594(0.012) & 13.068(0.009) & 12.774(0.012) &
12.475(0.011) & 12.02(0.05) & 11.85(0.03) & 11.77(0.03) \nl
2 (S7) & 15.437(0.029) & 15.551(0.015) & 15.016(0.013) & 14.715(0.015) & 
14.403(0.015) & \nodata & \nodata & \nodata \nl
3 (S1) & 15.523(0.035) & 15.520(0.018) & 14.895(0.017) & 14.553(0.019) & 
14.199(0.019) & \nodata & \nodata & \nodata \nl
4 (S2) & 16.501(0.040) & 16.495(0.027) & 15.789(0.024) & 15.384(0.026) & 
14.971(0.025) & \nodata & \nodata & \nodata \nl
5 (S8) & 16.778(0.039) & 16.280(0.025) & 15.441(0.024) & 14.987(0.027) & 
14.573(0.025) & \nodata & \nodata & \nodata \nl
6 (S12) & \nodata & 19.210(0.042) & 17.730(0.031) & 16.805(0.040) & 
15.583(0.033) & 14.36(0.05) & 13.75(0.04) & 13.54(0.03) \nl
\enddata
\tablecomments{The S identifiers are from Suntzeff et
al. (1999). Uncertainties in the magnitudes are listed in parentheses.}
\end{deluxetable}

To measure the brightness of SN 1998bu, we performed aperture
photometry of the supernova and comparison stars on each image. We
derived an aperture correction determined from one or a few isolated
bright stars measured through circular apertures of varying radii
(with sky measurements in a surrounding annulus). We were thus able to
measure the supernova light through a small aperture, so that noise
from the background sky would be minimized. Due to the varying seeing
and pixel scales at the sites, we did not impose a fixed angular size
aperture for all the observations. In all cases, though, the aperture
for the supernova instrumental magnitudes was the same size as the
aperture for the comparison stars in the field. We chose aperture
photometry over point-spread-function (PSF) fitting primarily out of
convenience, but also because in many instances the fields were too
small or the exposures too short to reliably determine the PSF from
nearby stars.

We also determined the linear color transformation coefficients for
each telescope/instrument/filter combination. At sites which had
photometric conditions, these were the same as those used in
determining comparison star magnitudes. For the other sites, color
terms were determined from observations of either Landolt fields or
other standard star fields but allowing for a varying zero point to
account for the non-photometric conditions. The transformation
coefficients were also checked with the local comparison stars. Color
terms for each telescope/filter combination are listed in Table
\ref{tabcolterm}. Uncertainties in the transformation coefficients
were propagated to the error estimate in the final photometry.

\begin{deluxetable}{cccccc}
\tablecaption{Photometric Color Terms \label{tabcolterm}}
\tablehead{\colhead{Telescope} & \colhead{$U-B$} & \colhead{$B-V$} & 
\colhead{$V-R$} & \colhead{$V-I$} & \colhead{$V (B-V)$}}
\startdata
CfA4 & 0.935 & 0.887 & 0.973 & 1.086 & +0.041 \nl
CfA1 & 0.926 & 0.939 & 0.982 & 1.068 & +0.027 \nl
MDM  & 0.782 & 1.061 & 0.981\tablenotemark{a} & 1.027 & +0.019 \nl
KP09 & 0.907 & 0.912 & 1.005 & 1.027 & +0.009 \nl
WIYN & 1.140 & 0.974 & 1.060 & 1.100 & +0.019 \nl
Gett & \nodata & 0.549\tablenotemark{b} & 0.894 & 1.006 & +0.009 \nl
Well & \nodata & 0.658\tablenotemark{b} & 0.980 & 1.099 & +0.030 \nl
KAIT & \nodata & 0.929 & 1.190 & 1.082 & +0.001 \nl
\enddata
\tablecomments{
The telescope designations are: CfA4, FLWO 1.2-m with 4Shooter; CfA1,
FLWO 1.2-m with AndyCam; MDM, MDM Hiltner 2.4-m; KP09, KPNO 0.9-m;
WIYN, WIYN 3.5-m; Gett, Gettysburg 0.4-m; Well, Wellesley 0.6-m; and
KAIT, KAIT 0.76-m. The tabulated values for the color columns ($U-B$,
$B-V$, $V-R$, $V-I$) are the transformation coeffiencents from the
standard color to the instrumental color, e.g., CfA4: $(u-b) =
0.935 (U-B) + {\rm const}$. The tabulated values for the last column
are transformation coefficients from $B-V$ to the instrumental $v$
magnitude, e.g., CfA4: $v-V = +0.041 (B-V) + {\rm const}$.}
\tablenotetext{a}{The MDM $R$-band photometry was not used; see text
for details.}
\tablenotetext{b}{As described in the text, we have not used a linear
transformation to place Gettysburg and Wellesley $B$-band supernova
photometry onto the standard system.}
\end{deluxetable}

Since our observations of SN 1998bu used local standard stars at the
same airmass as the supernova, no airmass correction was necessary for
the differential photometry. By using the measured color terms, we
only had to determine the zero point of each frame, by solving for the
offset between the comparison star color-corrected instrumental
magnitudes and their standard magnitudes. The zero point was
determined from the flux-weighted combination of all available
comparison stars. The scatter in the comparison star magnitudes
(typically $\sim 0.02$ mag) was used as an estimate of the internal
error in determining the zero point. From the derived zero point and
color transformation coefficient, we were then able to transform the
supernova instrumental magnitude into a standard magnitude.

This procedure was quite effective in the majority of cases. However,
supernova photometry requires special care. Where the color terms are
large (due to a mismatch between the filter/detector response and a
standard Landolt response), the use of a linear color correction can
be insufficient. This problem is especially acute in supernova
photometry because of deep, wide features in the spectra of
supernovae. Even when a large color correction works well when applied
to stars (i.e., yields very small differences with standard
magnitudes), such large corrections may not be appropriate to
supernovae. In most cases, filters were well-matched and the derived
color corrections were small, so nothing more complicated was
required. However, in two data sets (the Gettysburg and Wellesley $B$
filters), the color-corrected supernova magnitudes were significantly
discrepant with other data. In both of these cases the color terms
were quite large. To combat this problem, we determined a correction
based on the $B$ filter transmission and detector sensitivity
functions from these two sites. We used these response functions with
spectrophotometric observations of a number of SNe Ia at varying ages,
and determined magnitude corrections relative to the standard passband
defined by Bessell (1990). This procedure mirrors the use of
K-corrections for high-redshift supernovae
\markcite{kim96}\markcite{sch98}(Kim, Goobar, \& Perlmutter 1996;
Schmidt et al. 1998). Combining these corrections (usually $\sim 0.1$
to $0.2$ mag) with zero points determined from the comparison stars
brought these two data sets into good accord with the others, though
the photometry has signficantly higher uncertainty (typically $\sim
0.1$ mag).

In addition, we encountered a puzzle in the MDM $R$-band data, which
was discrepantly bright (by $\sim 0.1$ mag) compared to data from four
other telescopes at nearly the same epoch. However, in this case, the
derived $R$ color term was not very large. We were unable to procure
the filter and detector characteristics for this data set. Thus we
cannot provide an adequate explanation for this anomaly; it is
possible that the filter transmission is mismatched just such that a
color term derived from stellar observations (which were somewhat
limited in color) would be small (e.g., high transmission at some
wavelengths compensated by low transmission at others), while the
supernova spectrum at the epoch of the observations led to a
discrepant magnitude. Since four other telescopes provided mutually
consistent observations at the same epoch, we have disregarded the MDM
$R$ data. We were fortunate to possess contemporaneous observations so
that we could discover this discrepancy, and it illustrates the need
for a careful investigation of filter and detector characteristics
when combining observations of supernovae made at various sites
\markcite{sun99}\markcite{wel94}(Suntzeff et al. 1999; Wells et
al. 1994).

\begin{figure*}
\plotone{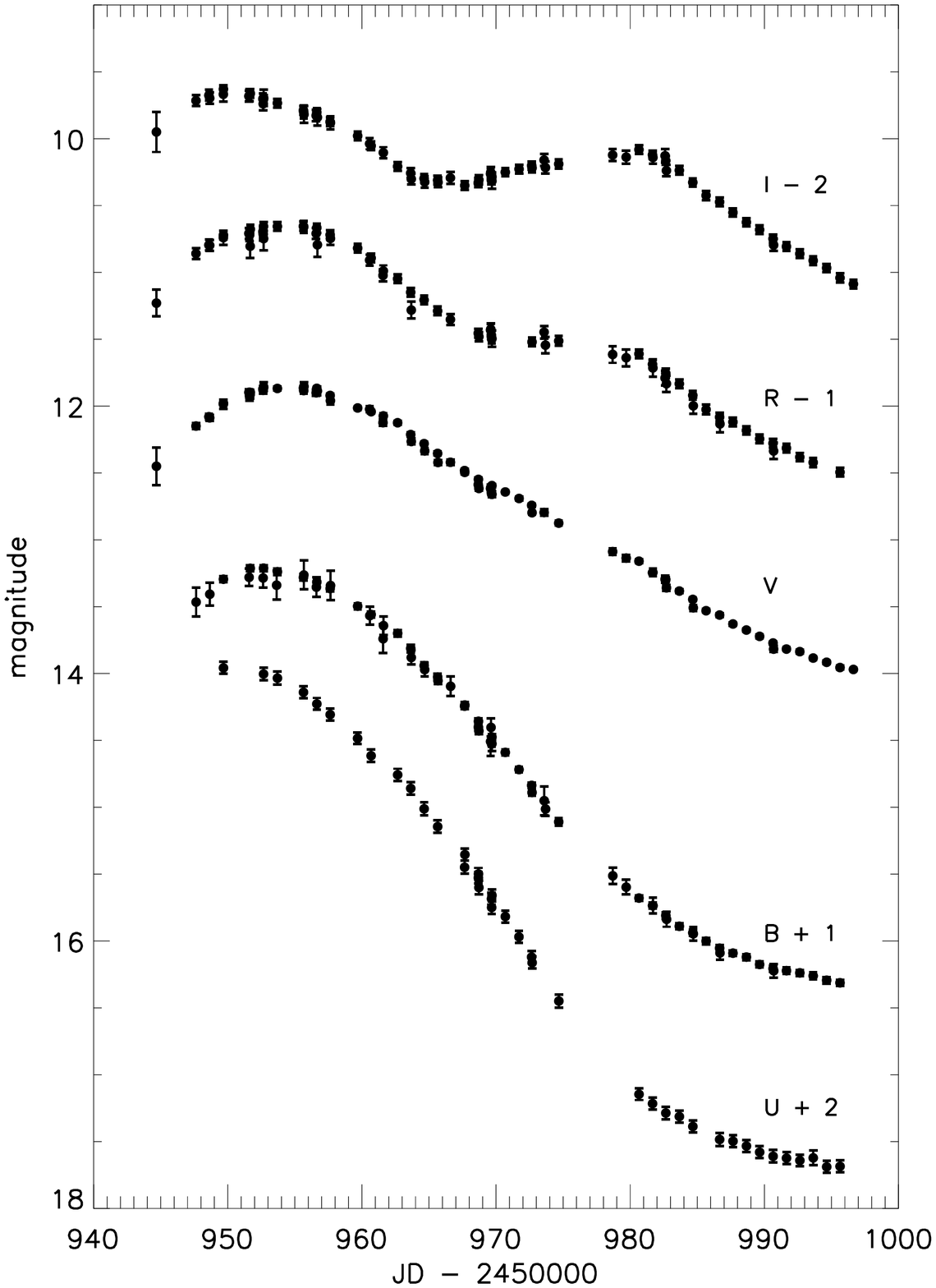}
\caption{$UBVRI$ light curves of SN 1998bu. \label{figophot}}
\end{figure*}

A futher complication in supernova photometry is proper discrimination
between light from the supernova and light from the underlying galaxy
\markcite{boi91}(Boisseau \& Wheeler 1991). Accurate subtraction of
the galaxy background is essential to measure correct magnitudes, and
even more important in measuring light curves, as a constant unremoved
galaxy background will cause a supernova light curve to look more
shallow (i.e., mimic a slower decline rate). Though SN 1998bu is
projected on a spiral arm of M96, the background from the galaxy is
relatively faint. We were fortunate to possess $UBVRI$ images of M96
prior to the appearance of SN 1998bu, taken with the FLWO 1.2-m for
another program. Our original plan was to use this image as a template
and apply the galaxy subtraction techniques described by Schmidt et
al. \markcite{sch98}\markcite{fil86}(1998; see also Filippenko et
al. 1986). This was very successful in a few cases, but it turned out
that the template image quality was much poorer (due to seeing and
the pixel scale) than
\begin{deluxetable}{ccccccc}
\tablecaption{$UBVRI$ Photometry of SN 1998bu. \label{tabophot}}
\tablehead{\colhead{Julian Day} & \colhead{$U$} & \colhead{$B$} & 
\colhead{$V$} & \colhead{$R$} & \colhead{$I$} & \colhead{Telescope}}
\startdata
2450944.68 & \nodata & \nodata & 12.45(0.14) & 12.23(0.10) &
11.95(0.15) & KAIT \nl 2450947.63 & \nodata & 12.47(0.11) &
12.15(0.02) & 11.86(0.04) & 11.72(0.04) & Gett \nl 2450948.59 &
\nodata & \nodata & 12.08(0.02) & 11.80(0.04) & 11.68(0.04) & Well \nl
2450948.65 & \nodata & 12.41(0.09) & 12.09(0.02) & 11.80(0.04) &
11.70(0.04) & Gett \nl 2450949.67 & \nodata & \nodata & 11.99(0.03) &
11.74(0.05) & 11.67(0.06) & Gett \nl 2450949.67 & 11.96(0.04) &
12.29(0.02) & 11.98(0.01) & 11.72(0.03) & 11.63(0.03) & CfA4 \nl
2450951.58 & \nodata & 12.28(0.07) & 11.90(0.02) & 11.71(0.04) &
11.68(0.04) & Well \nl 2450951.63 & \nodata & \nodata & 11.93(0.03) &
11.72(0.04) & 11.67(0.04) & Gett \nl 2450951.67 & \nodata & \nodata &
11.91(0.03) & 11.80(0.09) & \nodata & KAIT \nl 2450951.68 & \nodata &
12.21(0.02) & 11.90(0.01) & 11.68(0.03) & 11.66(0.03) & CfA4 \nl
2450952.62 & \nodata & 12.29(0.07) & 11.87(0.02) & 11.70(0.04) &
11.70(0.04) & Well \nl 2450952.64 & \nodata & \nodata & 11.88(0.02) &
11.71(0.05) & 11.74(0.05) & Gett \nl 2450952.66 & 12.00(0.05) &
12.21(0.02) & 11.86(0.01) & 11.66(0.03) & 11.71(0.03) & CfA4 \nl
2450952.67 & \nodata & \nodata & 11.85(0.03) & 11.75(0.09) &
11.69(0.06) & KAIT \nl 2450953.63 & \nodata & 12.34(0.11) & \nodata &
\nodata & \nodata & Gett \nl 2450953.70 & 12.03(0.05) & 12.24(0.02) &
11.87(0.01) & 11.66(0.03) & 11.73(0.03) & CfA4 \nl 2450955.64 &
12.14(0.05) & 12.28(0.02) & 11.86(0.01) & 11.65(0.03) & 11.79(0.03) &
CfA4 \nl 2450955.66 & \nodata & 12.26(0.11) & 11.88(0.02) &
11.66(0.04) & 11.80(0.05) & Gett \nl 2450955.68 & \nodata & \nodata &
11.85(0.03) & \nodata & 11.82(0.06) & KAIT \nl 2450956.59 & \nodata &
12.35(0.07) & 11.90(0.02) & 11.71(0.04) & 11.83(0.04) & Well \nl
2450956.64 & 12.23(0.04) & 12.32(0.02) & 11.87(0.01) & 11.67(0.03) &
11.80(0.03) & CfA4 \nl 2450956.68 & \nodata & \nodata & 11.89(0.03) &
11.79(0.09) & 11.84(0.06) & KAIT \nl 2450957.63 & 12.31(0.05) &
12.36(0.02) & 11.92(0.01) & 11.72(0.03) & 11.87(0.03) & CfA4 \nl
2450957.65 & \nodata & 12.34(0.11) & 11.96(0.03) & 11.75(0.05) &
11.88(0.05) & Gett \nl 2450959.68 & 12.48(0.04) & 12.50(0.02) &
12.01(0.01) & 11.82(0.03) & 11.98(0.03) & CfA1 \nl 2450960.57 &
\nodata & 12.57(0.07) & 12.02(0.02) & 11.91(0.04) & 12.04(0.04) & Well
\nl 2450960.69 & 12.61(0.05) & 12.56(0.02) & 12.04(0.01) & 11.89(0.03)
& 12.05(0.03) & CfA1 \nl 2450961.57 & \nodata & 12.74(0.11) &
12.12(0.03) & 12.02(0.05) & \nodata & Gett \nl 2450961.60 & \nodata &
12.64(0.07) & 12.07(0.02) & 11.99(0.04) & 12.10(0.04) & Well \nl
2450962.66 & 12.76(0.05) & 12.70(0.02) & 12.12(0.01) & 12.05(0.03) &
12.21(0.03) & CfA4 \nl 2450963.64 & 12.86(0.05) & 12.81(0.03) &
12.21(0.02) & 12.15(0.03) & 12.26(0.04) & CfA4 \nl 2450963.68 &
\nodata & 12.88(0.05) & 12.26(0.02) & 12.28(0.06) & 12.30(0.04) & KAIT
\nl 2450964.63 & 13.01(0.05) & 12.94(0.03) & 12.28(0.01) & 12.21(0.03)
& 12.30(0.03) & CfA4 \nl 2450964.68 & \nodata & 12.97(0.05) &
12.34(0.02) & \nodata & 12.32(0.04) & KAIT \nl 2450965.64 &
13.14(0.05) & 13.03(0.02) & 12.35(0.01) & 12.29(0.03) & 12.31(0.03) &
CfA4 \nl 2450965.67 & \nodata & 13.05(0.03) & 12.42(0.02) & \nodata &
12.33(0.04) & WIYN \nl 2450966.60 & \nodata & 13.10(0.07) &
12.42(0.02) & 12.35(0.04) & 12.29(0.04) & Well \nl 2450967.65 &
13.45(0.05) & 13.24(0.02) & 12.48(0.01) & \nodata & \nodata & CfA4 \nl
2450967.66 & 13.35(0.05) & 13.24(0.02) & 12.50(0.01) & \nodata &
12.35(0.03) & MDM \nl 2450968.67 & 13.53(0.04) & 13.40(0.02) &
12.59(0.01) & 12.46(0.03) & 12.33(0.03) & KP09 \nl 2450968.68 &
13.50(0.04) & 13.36(0.02) & 12.55(0.01) & \nodata & 12.33(0.03) & MDM
\nl 2450968.72 & 13.60(0.05) & 13.43(0.03) & 12.62(0.02) & 12.48(0.03)
& 12.31(0.03) & WIYN \nl 2450969.59 & \nodata & 13.51(0.11) &
12.62(0.02) & 12.43(0.05) & 12.26(0.04) & Gett \nl 2450969.60 &
\nodata & 13.40(0.07) & 12.61(0.02) & 12.43(0.04) & 12.25(0.04) & Well
\nl 2450969.65 & 13.68(0.04) & 13.51(0.02) & 12.64(0.01) & 12.47(0.03)
& 12.31(0.03) & KP09 \nl 2450969.67 & 13.75(0.05) & 13.52(0.03) &
12.66(0.02) & 12.50(0.03) & 12.30(0.03) & WIYN \nl 2450969.69 &
\nodata & 13.53(0.05) & 12.66(0.02) & 12.50(0.06) & 12.32(0.06) & KAIT
\nl 2450969.69 & 13.66(0.04) & 13.47(0.02) & 12.59(0.01) & \nodata &
12.28(0.03) & MDM \nl 2450970.69 & 13.82(0.05) & 13.59(0.02) &
12.64(0.01) & \nodata & 12.25(0.03) & MDM \nl 2450971.71 & 13.97(0.05)
& 13.72(0.02) & 12.69(0.01) & \nodata & 12.23(0.03) & MDM \nl
2450972.65 & 14.12(0.04) & 13.84(0.02) & 12.74(0.01) & \nodata &
12.20(0.03) & MDM \nl 2450972.68 & 14.16(0.04) & 13.89(0.02) &
12.80(0.01) & 12.52(0.03) & 12.22(0.03) & WIYN \nl 2450973.58 &
\nodata & 13.95(0.11) & 12.80(0.02) & 12.45(0.05) & 12.16(0.04) & Gett
\nl 2450973.68 & \nodata & 14.01(0.05) & \nodata & 12.54(0.06) &
12.22(0.04) & KAIT \nl 2450974.67 & 14.45(0.05) & 14.11(0.03) &
12.87(0.02) & 12.51(0.04) & 12.19(0.04) & WIYN \nl 2450978.69 &
\nodata & 14.51(0.06) & 13.09(0.02) & 12.61(0.06) & 12.12(0.05) & KAIT
\nl 2450979.70 & \nodata & 14.60(0.06) & 13.14(0.03) & 12.64(0.06) &
12.14(0.05) & KAIT \nl 2450980.65 & 15.15(0.04) & 14.68(0.02) &
13.16(0.01) & 12.61(0.03) & 12.08(0.03) & CfA4 \nl 2450981.66 &
15.22(0.04) & 14.74(0.02) & 13.25(0.01) & 12.69(0.03) & 12.12(0.03) &
CfA4 \nl 2450981.69 & \nodata & 14.74(0.06) & 13.24(0.03) &
12.71(0.07) & 12.14(0.05) & KAIT \nl 2450982.60 & \nodata & \nodata &
13.30(0.03) & 12.79(0.06) & 12.13(0.05) & Gett \nl 2450982.65 &
15.29(0.05) & 14.81(0.02) & 13.30(0.02) & 12.75(0.03) & 12.16(0.03) &
CfA4 \nl 2450982.70 & \nodata & 14.84(0.06) & 13.36(0.02) &
12.83(0.06) & 12.24(0.04) & KAIT \nl 2450983.66 & 15.31(0.05) &
14.89(0.02) & 13.38(0.01) & 12.83(0.03) & 12.24(0.03) & CfA4 \nl
2450984.67 & 15.39(0.04) & 14.94(0.02) & 13.45(0.01) & 12.92(0.03) &
12.33(0.03) & CfA4 \nl 2450984.70 & \nodata & 14.95(0.05) &
13.51(0.02) & 13.00(0.06) & \nodata & KAIT \nl 2450985.65 & \nodata &
15.00(0.02) & 13.53(0.01) & 13.03(0.04) & 12.42(0.03) & CfA4 \nl
2450986.66 & 15.48(0.05) & 15.05(0.02) & 13.56(0.01) & 13.08(0.03) &
12.47(0.03) & CfA4 \nl 2450986.69 & \nodata & 15.09(0.05) & \nodata &
13.13(0.06) & \nodata & KAIT \nl 2450987.66 & 15.50(0.04) &
15.09(0.02) & 13.63(0.01) & 13.12(0.03) & 12.55(0.03) & CfA4 \nl
2450988.66 & 15.53(0.04) & 15.12(0.02) & 13.67(0.01) & 13.18(0.03) &
12.63(0.03) & CfA4 \nl 2450989.65 & 15.58(0.04) & 15.18(0.02) &
13.72(0.01) & 13.24(0.03) & 12.68(0.03) & CfA4 \nl 2450990.65 &
15.61(0.05) & 15.21(0.02) & 13.77(0.01) & 13.28(0.03) & 12.75(0.03) &
CfA4 \nl 2450990.69 & \nodata & 15.22(0.05) & 13.82(0.02) &
13.34(0.06) & 12.80(0.05) & KAIT \nl 2450991.65 & 15.62(0.05) &
15.22(0.02) & 13.82(0.01) & 13.31(0.03) & 12.81(0.03) & CfA4 \nl
2450992.64 & 15.64(0.04) & 15.24(0.02) & 13.84(0.01) & 13.38(0.03) &
12.86(0.03) & CfA4 \nl 2450993.64 & 15.62(0.06) & 15.26(0.03) &
13.88(0.02) & 13.42(0.03) & 12.91(0.03) & CfA4 \nl 2450994.64 &
15.69(0.05) & 15.29(0.02) & 13.92(0.01) & \nodata & 12.97(0.03) & CfA4
\nl 2450995.64 & 15.69(0.04) & 15.31(0.02) & 13.96(0.01) & 13.49(0.03)
& 13.04(0.03) & CfA4 \nl 2450996.64 & \nodata & \nodata & 13.97(0.01)
& \nodata & 13.09(0.03) & CfA4 \nl
\enddata
\tablecomments{The telescope designations are as in Table \ref{tabcolterm}.}
\end{deluxetable} 
\noindent many of our observations, and degrading the
supernova observations to match the template added undesirable
correlated noise. In additional cases the observed fields did not
align well with the template image (because there were very few stars
in some of the small-field observations). Instead we estimated the
galaxy background flux directly from the template image (using an
aperture and sky annulus appropriate to each observation) and
subtracted this flux from that of the supernova. The correction to the
supernova magnitude was initially negligibly small and grew larger as
the supernova faded, but the maximum correction made was only 0.025
mag. Our prediscovery images justify the assumption of a small
host-galaxy flux contribution made by \markcite{sun99}Suntzeff et
al. (1999). Our final photometry for SN 1998bu is listed in Table
\ref{tabophot} and the optical light curves are shown in Figure
\ref{figophot}.

\subsection{Infrared Photometry}
 
Infrared photometry of SN~1998bu in the $JHK$ passbands was obtained
at the Fred L. Whipple Observatory (FLWO), the Mount Stromlo
Observatory (MSO), the Infrared Telescope Facility (IRTF), and with
the ESO New Technology Telescope (NTT).  The FLWO data were obtained
with the 1.2-m telescope and ``STELIRCam'' IR camera which consists of
two 256$^2$ InSb detector arrays permitting simultaneous imaging in
two filters \markcite{tol99}(Tollestrup et al. 1999). The FLWO filters
were manufactured by Barr Associates in 1987 for a number of
institutions including NOAO. The MSO data was taken with the 2.3-m
telescope and CASPIR Spectrograph/Imager which uses a 256$^2$ InSb
detector. The IRTF data was obtained with the NASA 3-m telescope and
NSFCAM IR camera which uses a 256$^2$ InSb detector. The NTT is a
3.5-m aperture telescope and observations were made using the SOFI
imaging spectrograph which employs a 1024$^2$ HgCdTe array.

\begin{figure*}
\plotone{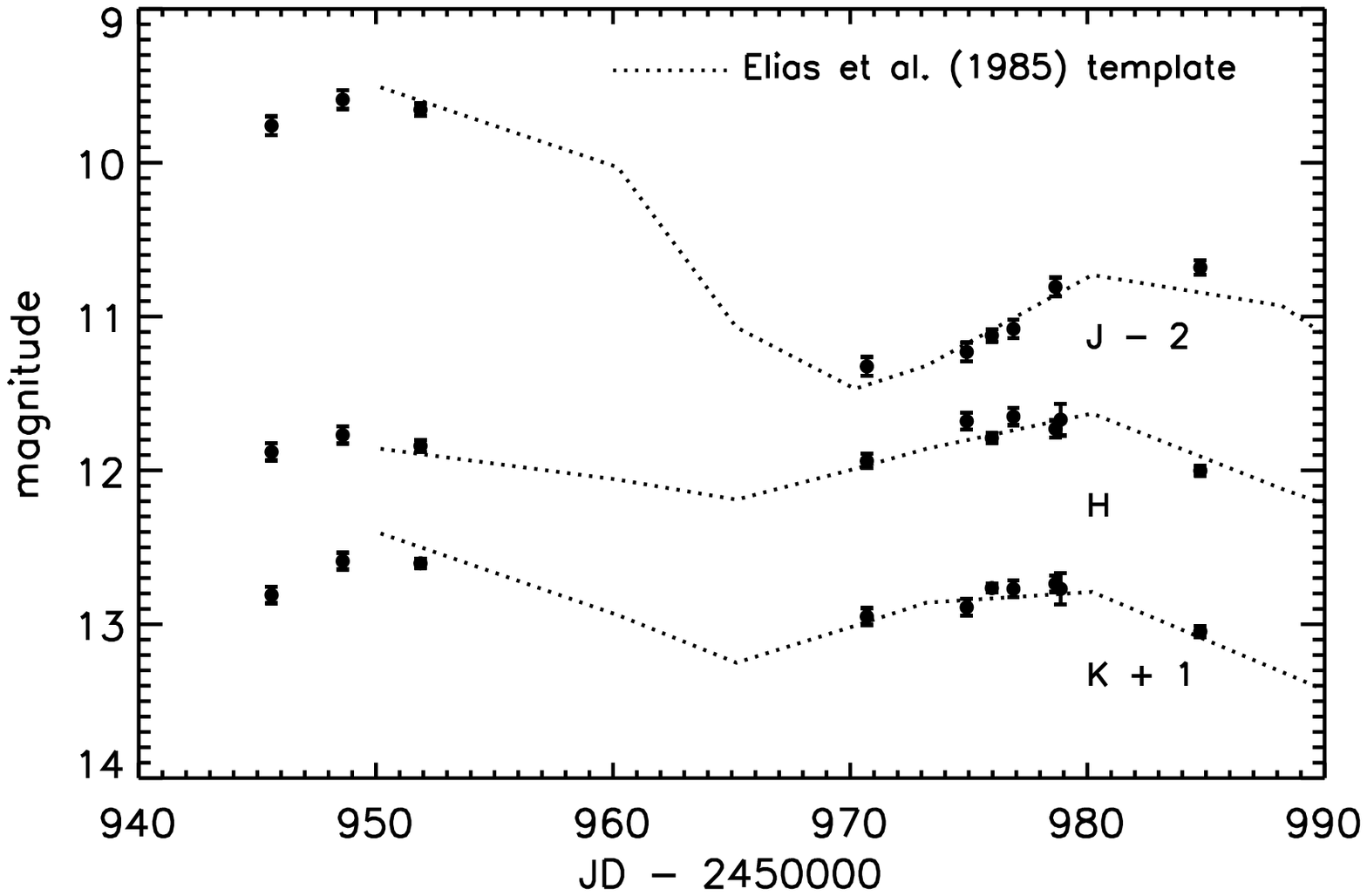} 
\caption{$JHK$ light curves of SN 1998bu. \label{figirphot}}
\end{figure*}

Flat fields and sky frames were created using offset field images
staggered between the supernova exposures. The alternating offset
field frames were subtracted from the corresponding data images
and then divided by the normalized, average flat field. Standard stars
for the FLWO observations were taken from \markcite{eli82}Elias et
al. (1982), while the NTT used standards from \markcite{pers98}Persson
et al. (1998), and the IRTF used the UKIRT faint standards
\markcite{hunt98}(Hunt et al. 1998). The MSO data was calibrated with
standards from \markcite{cart95}Carter \& Meadows (1995). Where
possible, the results were transformed to the Elias system, but the
variety of detectors, filters and standards introduces systematic
errors on the order of 0.05 mag. Two stars (stars 1 and 6) near the
supernova were calibrated from the FLWO data and used as secondary
standards on non-photometric nights; their magnitudes are listed in
Table \ref{tabcstar}. The galaxy background at the position of the
supernova is smooth and much fainter than the supernova light, so that
aperture photometry was sufficient. The resulting supernova photometry
is given in Table \ref{tabirphot} and displayed in Figure
\ref{figirphot}.

\begin{deluxetable}{ccccc}
\tablecaption{$JHK$ Photometry of SN 1998bu. \label{tabirphot}}
\tablehead{\colhead{Julian Day} & \colhead{$J$} & \colhead{$H$} & 
\colhead{$K$} & \colhead{Telescope}}
\startdata
2450945.6\phn  & 11.76(0.06) & 11.88(0.06) & 11.81(0.05) & FLWO   \nl
2450948.6\phn  & 11.59(0.06) & 11.77(0.06) & 11.59(0.05) & FLWO   \nl
2450951.88     & 11.66(0.04) & 11.84(0.04) & 11.60(0.03) & IRTF   \nl
2450970.7\phn  & 13.32(0.06) & 11.94(0.05) & 11.95(0.05) & FLWO   \nl
2450974.91     & 13.23(0.06) & 11.68(0.06) & 11.89(0.05) & MSO    \nl
2450975.97     & 13.12(0.04) & 11.79(0.03) & 11.77(0.03) & ESO    \nl
2450976.88     & 13.08(0.06) & 11.65(0.06) & 11.77(0.05) & MSO    \nl
2450978.65     & 12.81(0.06) & 11.73(0.06) & 11.74(0.05) & FLWO   \nl
2450978.87     & \nodata     & 11.67(0.10) & 11.77(0.10) & MSO    \nl
2450984.76     & 12.68(0.05) & 12.00(0.03) & 12.05(0.04) & IRTF   \nl
\enddata
\end{deluxetable}

\subsection{Optical Spectroscopy}

All of our optical spectroscopic observations were obtained at the
FLWO 1.5-m telescope with the FAST spectrograph. This long-slit
spectrograph has been designed for high throughput and features a
thinned, back-side illuminated, anti-reflection coated CCD
detector. The slit length is $180\arcsec$ and can be adjusted to
several widths; we have generally employed a $3\arcsec$ slit for
our observations. These were made using a 300 line/mm grating which
results in a resolution of roughly 0.6 nm and a usable wavelength range
from 360 to 720 nm. We began observations of SN 1998bu on May 16.1 and
continued through July 15.1, on a total of 27 nights.

\begin{figure*}
\plotone{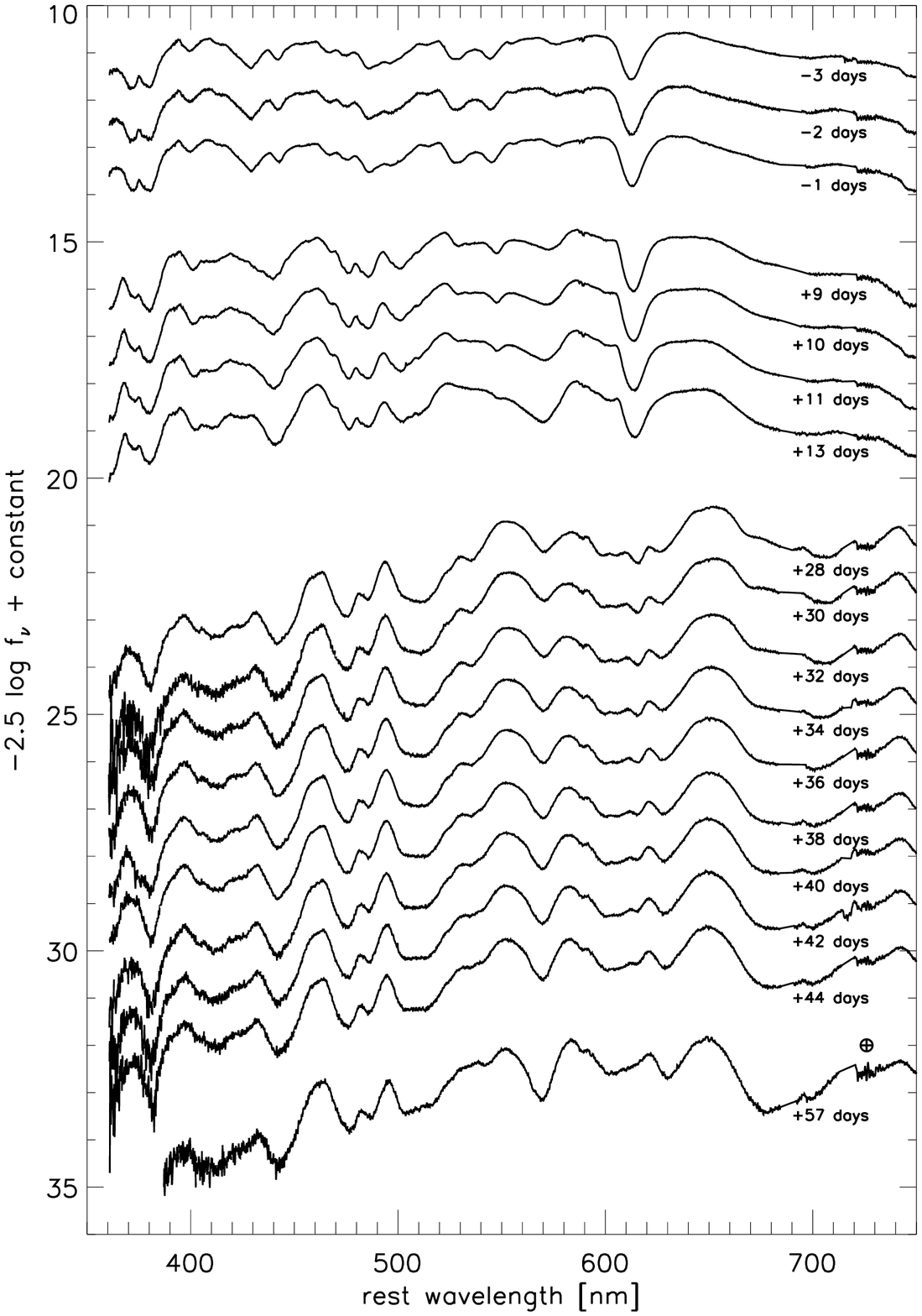}
\caption{Optical spectra of SN 1998bu labelled by epoch with respect
to $B$ maximum. For clarity the spectra have been shifted vertically
by arbitrary amounts. Unremoved telluric features are
marked.\label{figspcevol}}
\end{figure*}

We have reduced the spectra in the standard manner with IRAF. The
two-dimensional CCD exposures were corrected for bias and dark-current
and were flat-fielded using CCDPROC. A 1D spectrum was extracted at
the supernova position subtracting the neigboring sky using the
APEXTRACT task. Wavelength calibration was performed by extracting the
same aperture from an exposure of a HeNeAr lamp taken just after the
supernova observation and identifying emission lines. We also
performed flux calibration through the reduction of a
spectrophotometric standard star each night \markcite{mas88}(Massey et
al. 1988). The conditions for the supernova observations were usually
not photometric, so the absolute flux calibration is generally
unreliable. The relative flux measurement may also suffer because of
differential refraction \markcite{fil82}(Filippenko 1982), as the slit
was always oriented east-west (PA $= 90\arcdeg$) rather than being
optimally oriented at the parallactic angle. In general, then, our
spectra likely underestimate the flux in the blue by 10 to 20\%. In
addition, since we have not used a blocking filter, second-order light
contaminates the red end of the spectra. The supernova flux in the
blue is generally lower than in the red and the detector sensitivity
to the blue photons is also low, so that the second-order
contamination is on the order of only a few percent. Second order
contamination in the standard star spectra is more significant (since
the standard stars are typically quite blue), so that the flux
calibration in the red is also somewhat uncertain. Uncertainties in
the flux calibration in both the blue and the red thus make these
spectra unsuitable for spectrophotometry. Contamination of the
supernova spectra by underlying galaxy light (after subtracting the
local sky) was small, as in the case of the optical photometry. In
some cases, multiple observations on the same night were combined into
one. Cosmic rays and telluric lines were removed by hand. Figure
\ref{figspcevol} shows a representative subset of our optical
spectroscopy of SN 1998bu and the spectral evolution of the supernova
covering 60 days, from approximately 3 days before maximum light in
the $B$ band. The complete spectroscopic data set is available upon
request.

\subsection{Infrared Spectroscopy}

Infrared spectra of SN~1998bu were obtained with the KPNO 4-m
telescope and OSU-NOAO Infrared Imager (ONIS) on 1998 June 14.2
UT. The ONIS covered the $K$-band range from 2.0 ${\mu}$m to 2.4
${\mu}$m with a resolution of 1.6 nm (FWHM). A sequence of two-minute
exposures was made, while stepping the target along the slit. These
were combined for a total integration of 24 minutes. A spectrum of the
F5V star BS~4281 was divided into the supernova spectrum to remove
telluric absorption (except in the deep absorption bands at 1.4 and
1.9 $\mu$m). A smooth spectrum of an F5V star was created by
interpolating the broad band colors \markcite{joh68}(Johnson et
al. 1968), with the zero point set by the catalog magnitude of
BS~4281. Multiplying the supernova spectrum and the synthetic F5V star
spectrum corrects for the detector sensitivity variations and roughly
calibrates supernova flux. The spectral flux was then adjusted to match
the observed $K$-band photometric magnitude of the supernova interpolated
to the date of the spectrum.
 
A spectrum was also taken with the SOFI instrument on the ESO NTT on
1998 June 11.0 UT. SOFI covered 0.95 ${\mu}$m to 2.5 ${\mu}$m in two
grating settings with significant overlap. Four 120-s exposures were
obtained at each tilt at four slit positions allowing good sky
subtraction. Spectra of HD177619 (an F7V star) were used to remove
telluric bands and calibrate the relative sensitivity of the detector
as described above. Absolute flux calibration was done using the
$H$-band magnitude determined from SOFI imaging done on the same night.
 
\begin{figure*}
\plotone{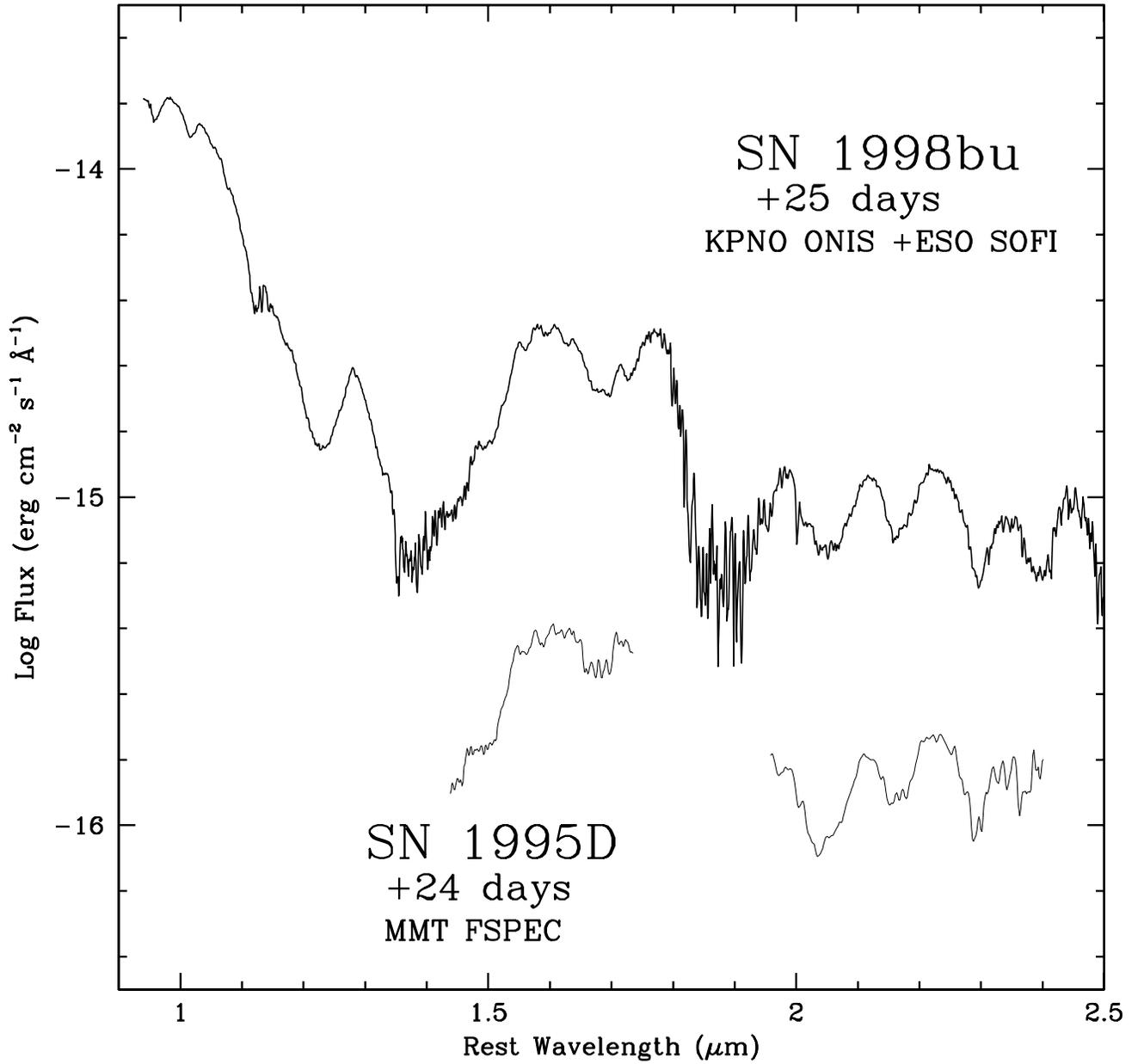}
\caption{Composite near-infrared spectrum of SN 1998bu at
approximately 25 days past maximum light. A spectrum of the type Ia SN
1995D at roughly the same epoch is shown for
comparison. \label{figirspc}}
\end{figure*}

The KPNO and ESO spectra were combined into a single high-quality
spectrum of SN~1998bu at an age of about +25 days which is shown in
Figure \ref{figirspc}. There are few good infrared spectra of SNe Ia
at the same epoch to compare with these data; however, a spectrum of
SN 1995D taken with the MMT+FSPEC at an age of +24 days is also shown
for comparison.

\section{Results}

\subsection{Spectra}

\begin{figure*}
\plotone{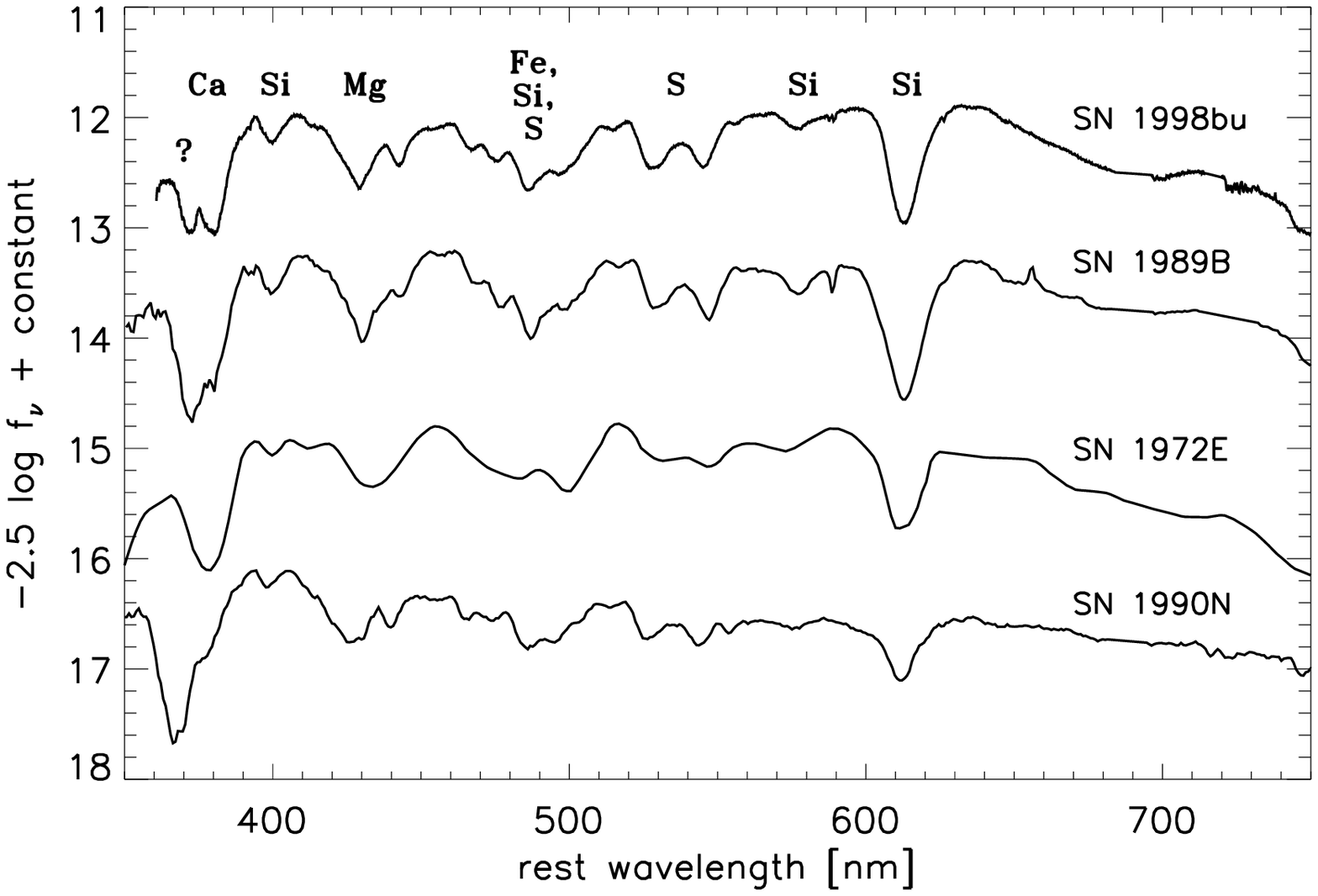}
\caption{Optical spectra near maximum light for SN 1998bu and
prototypical SNe Ia: SN 1989B (Wells et al. 1994), SN 1972E (Kirshner
et al. 1973; observed at significantly lower resolution than the
others), and SN 1990N (Leibundgut et al. 1991a). The spectra show
remarkable homogeneity and place SN 1998bu squarely among the typical
SNe Ia. \label{figmaxspc}}
\end{figure*}

Type I supernovae are distinguished from those of type II by the
absence of hydrogen in their spectra, and SNe Ia are further
distinguished from SNe Ib and SNe Ic by the prominent \ion{Si}{2}
($\lambda$635.5 nm) absorption at maximum light (for examples and a
detailed discussion see \markcite{fil97}Filippenko 1997). Other
absorption features in the optical at maximum light are predominantly
from intermediate mass elements (Si, Ca, S, O, Mg) at high
velocity ($\sim 10,000 \; {\rm km \; s^{-1}}$) in the outer layers of
the supernova ejecta. Lines of Fe become prominent at about two weeks
after maximum light as the effective photosphere recedes into the
ejecta, and about a month after maximum light the supernova enters the
optically-thin nebular phase where forbidden emission lines of
iron-peak elements (Fe, Co) dominate \markcite{kir75}(Kirshner \& Oke
1975). The optical spectra of SN 1998bu shown in Figure
\ref{figspcevol} follow this typical evolution. A more detailed
comparison is illustrated in Figure \ref{figmaxspc} where we show the
optical spectra of SN 1998bu and other prototypical SNe Ia near
maximum light. The spectra are remarkably similar, though there are
some differences in the detailed shapes and velocities of the
features.

\begin{figure*}
\plotone{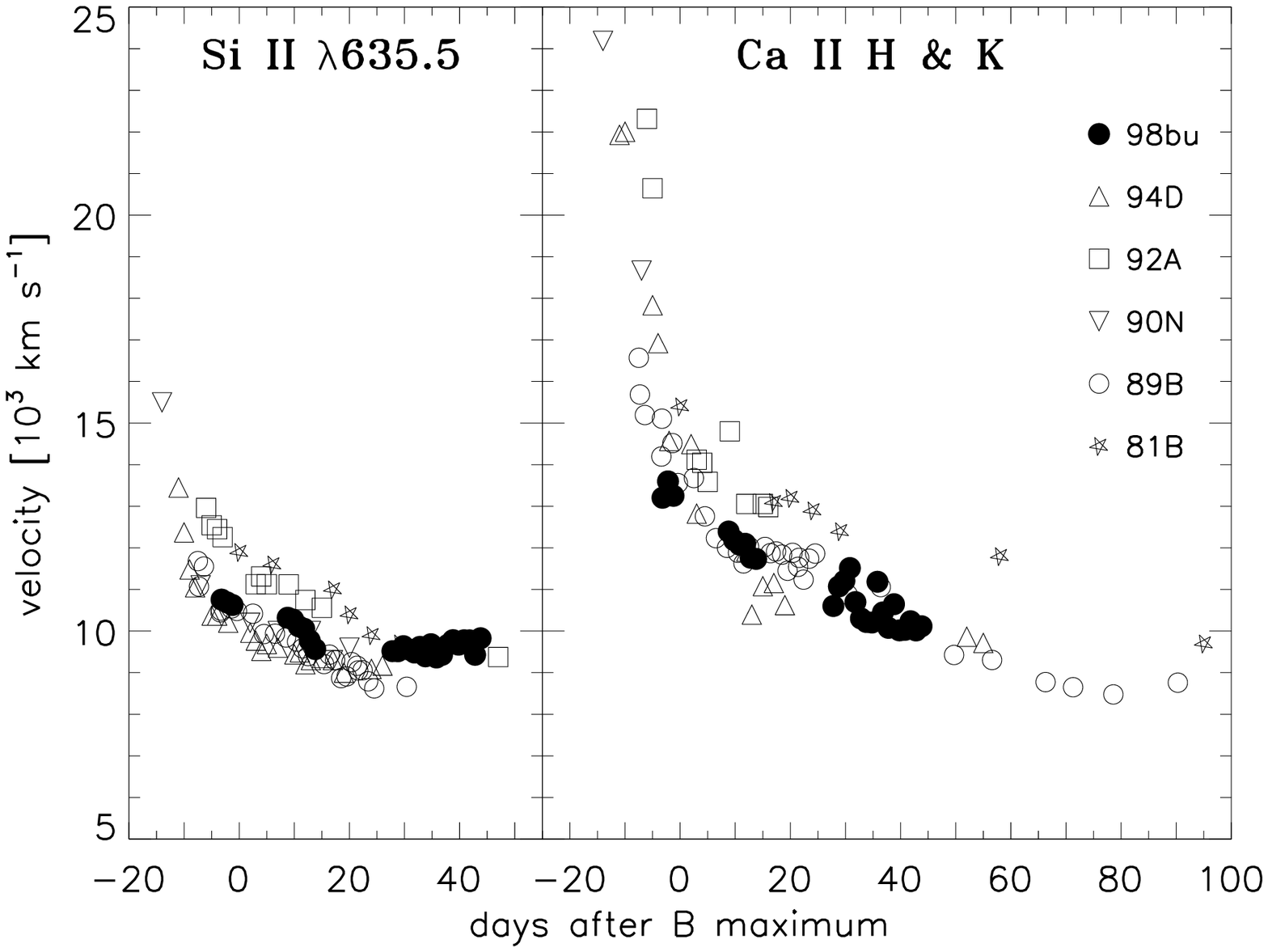}
\caption{Expansion velocities derived from the absorption minima of
\ion{Si}{2} $\lambda 635.5$~nm and \ion{Ca}{2} H \& K for SN 1998bu
(\emph{filled circles}), SN 1994D (\emph{triangles}), SN 1992A
(\emph{squares}), SN 1990N (\emph{upside-down triangles}), SN 1989B
(\emph{open circles}), and SN 1981B (\emph{stars}). The expansion
velocities have been corrected for the recession velocity of the host
galaxy. See text for references.\label{figvel}}
\end{figure*}

A more quantitative comparison between SN 1998bu and other typical SNe
Ia is illustrated in Figure \ref{figvel}. We show the velocities of
\ion{Si}{2} $\lambda 635.5$~nm and \ion{Ca}{2} H \& K flux minima as a
function of supernova phase for a number of prototypical SNe Ia: SN
1994D \markcite{pat96}(Patat et al. 1996), SN 1992A
\markcite{kir93}(Kirshner et al. 1993), SN 1990N
\markcite{lei91a}(Leibundgut et al. 1991a), SN 1989B
\markcite{bar90}\markcite{wel94}(Barbon et al. 1990; Wells et
al. 1994), and SN 1981B \markcite{bra83}(Branch et al. 1983).  SN
1998bu falls well within the scatter defined by these other objects.

However, the measurement of \ion{Ca}{2} H \& K velocities is made more
challenging by the presence of a feature blueward of the Ca feature
(indicated in Figure \ref{figmaxspc} by a question mark), which may be
due to either Si or possibly high-velocity Ca \markcite{hat99}(Branch
1998, personal communication; Hatano et al. 1999). For SN 1998bu this
unidentified feature was well separated from the normal Ca feature, as
is seen in the early spectra of the sequence in Figure
\ref{figspcevol}. The feature weakens with time such that by day +28
only the normal Ca feature remains. The sequence suggests that the Ca
absorption velocity should be defined by the redder of the two
troughs. For other SNe Ia, Figure \ref{figmaxspc} shows that the
unidentified feature and the Ca feature are not always well separated,
such as in the SN 1989B and SN 1990N spectra, where only a shoulder is
visible rather than two distinct minima. The spectrum of SN 1972E was
taken at significantly lower resolution, and there it looks like a
single absorption feature. Thus comparing measured \ion{Ca}{2} H \& K
velocities is tricky; our use of the red trough may lead to a
systematically lower velocity measured at early times, as seems to be
the case in Figure \ref{figvel}.

Supernovae with unusual luminosity also have unusual spectra.  Both SN
1991T, one of the most luminous SNe Ia known, and SN 1991bg, one of
the least luminous, showed spectral peculiarities in their maximum
light spectra. SN 1991T lacked a well-defined \ion{Si}{2} $\lambda
635.5$~nm feature at maximum
\markcite{fil92a}\markcite{phi92}(Filippenko et al. 1992a; Phillips et
al. 1992), though its subsequent evolution was similar to other SNe
Ia. On the other hand, maximum light spectra of SN 1991bg showed a
deep absorption trough attributed to \ion{Ti}{2} centered at a
wavelength of 420 nm \markcite{fil92b}\markcite{lei93}(Filippenko et
al. 1992b; Leibundgut et al. 1993). The absence of spectroscopic
peculiarities in SN 1998bu makes it a suitable calibrator of the SN Ia
distance scale.

The infrared spectrum of SN~1998bu in Figure \ref{figirspc} is one of
the best obtained for a type Ia supernova. It is strikingly similar to
an IR spectrum of SN~1995D taken at the same phase. SN 1995D also
showed no spectroscopic peculiarities in the optical and a typical
light curve \markcite{rie99}(Riess et al. 1999), bolstering the
evidence that SN~1998bu is a fairly normal SN Ia. The IR spectra are
qualitatively a good match to the model spectra of
\markcite{whe98}Wheeler et al. (1998), although the models do not
extend to the observed age of SN~1998bu and are meant to fit the
peculiar event SN~1986G. In the $K$-band ($\sim 2.2~{\mu}$m), which
consists of absorption features of Co, Ni, and Si, the spectra of SNe
Ia 1998bu, 1995D and 1986G \markcite{whe98}\markcite{fro87}(Wheeler et
al. 1998; Frogel et al. 1987) are nearly identical from 14 days past
maximum onwards. The $H$-band ($\sim 1.6~{\mu}$m) spectra of SN~1998bu
and SN~1995D do not show as large a gap between the peaks at 1.6
${\mu}$m and 1.8 ${\mu}$m as does SN~1986G. The deficit at 1.7
${\mu}$m is not visible at all in the peculiar SN~1991T
\markcite{bow97}(Bowers et al. 1997), but that spectrum was taken at a
more advanced age than the others. These observations suggest that the
1.7 ${\mu}$m gap depth may possibly be correlated with light curve
decline rate (and therefore luminosity) and supports the idea of
\markcite{whe98}Wheeler et al. (1998) that the 1.7 ${\mu}$m gap is an
indicator of the highest velocity of the Ni/Co region, though clearly
more infrared spectra of SNe Ia are required to test this
hypothesis. As in the optical spectra, the infrared spectra do not
indicate that SN 1998bu was peculiar.

\subsection{Light Curves and Peak Brightness}

The optical light curves of SN 1998bu presented in Figure
\ref{figophot} are among the best sampled early-time light curves of
any SN Ia. The $U$ band observations are particularly valuable as SNe
Ia have not often been observed in this passband. In addition, our
observations began before maximum light (unless stated otherwise, we
take maximum light to mean the time of maximum brightness in the $B$
band). These light curves are typical of SNe
Ia; the $UBV$ curves are well fit by the templates of
\markcite{lei89}Leibundgut (1989). In section \ref{sectmlcs} we
describe the detailed analysis of these light curve shapes, an
essential part of using this SN Ia as a distance indicator.

Work on SNe Ia as standard or calibrated candles employs the maximum
brightness in the $B$ and $V$ bands. We have determined the peak
apparent magnitude and time of maximum in the $B$-band for SN 1998bu
using a simple quadratic fit to the points within roughly five days of
the light curve peak, weighted by their photometric uncertainties. The
results are
\begin{equation}
{\rm JD_{Bmax}} = 2450952.8 \pm 0.8{\rm ,}
\end{equation}
\begin{equation}
B_{\rm Bmax} = 12.22 \pm 0.03.
\end{equation}
(Throughout the paper, all apparent and absolute magnitudes are
expressed in units of mag). The supernova apparent magnitude in $V$
\emph{at the time of $B$ maximum given above} is
\begin{equation}
V_{\rm Bmax} = 11.88 \pm 0.02.
\end{equation}

The time when the supernova was brightest in the $V$ band was JD
$2450954.4 \pm 1.0$, at $V = 11.86 \pm 0.02$. These results are quite
consistent with the finding of \markcite{lei89}Leibundgut (1989) that
maximum light in $V$ occurs about two days after maximum light in $B$
and his result that the $V$ magnitude at that time is $0.02$ mag
brighter than the $V$ magnitude at the time of $B$ maximum.  The time
of maximum brightness in the $U$ band is not well determined because
of the starting point of our data set, but our data are consistent
with the \markcite{lei89}Leibundgut (1989) result of $U$ maximum
occurring three days before maximum light in $B$. The $U$ apparent
magnitude at $B$ maximum is $U_{\rm Bmax} = 12.01 \pm 0.05$. SNe Ia
show increased variety in their $R$ and $I$ light curves as compared
to the bluer passbands, illustrated clearly in the composite light
curves presented by \markcite{rie99}Riess et al. (1999). The $R$ and
$I$ light curves of SN 1998bu are consistent with these composite
curves both in their general shape and times of maximum in those
bands. 

\begin{figure*}
\epsscale{0.90}
\plotone{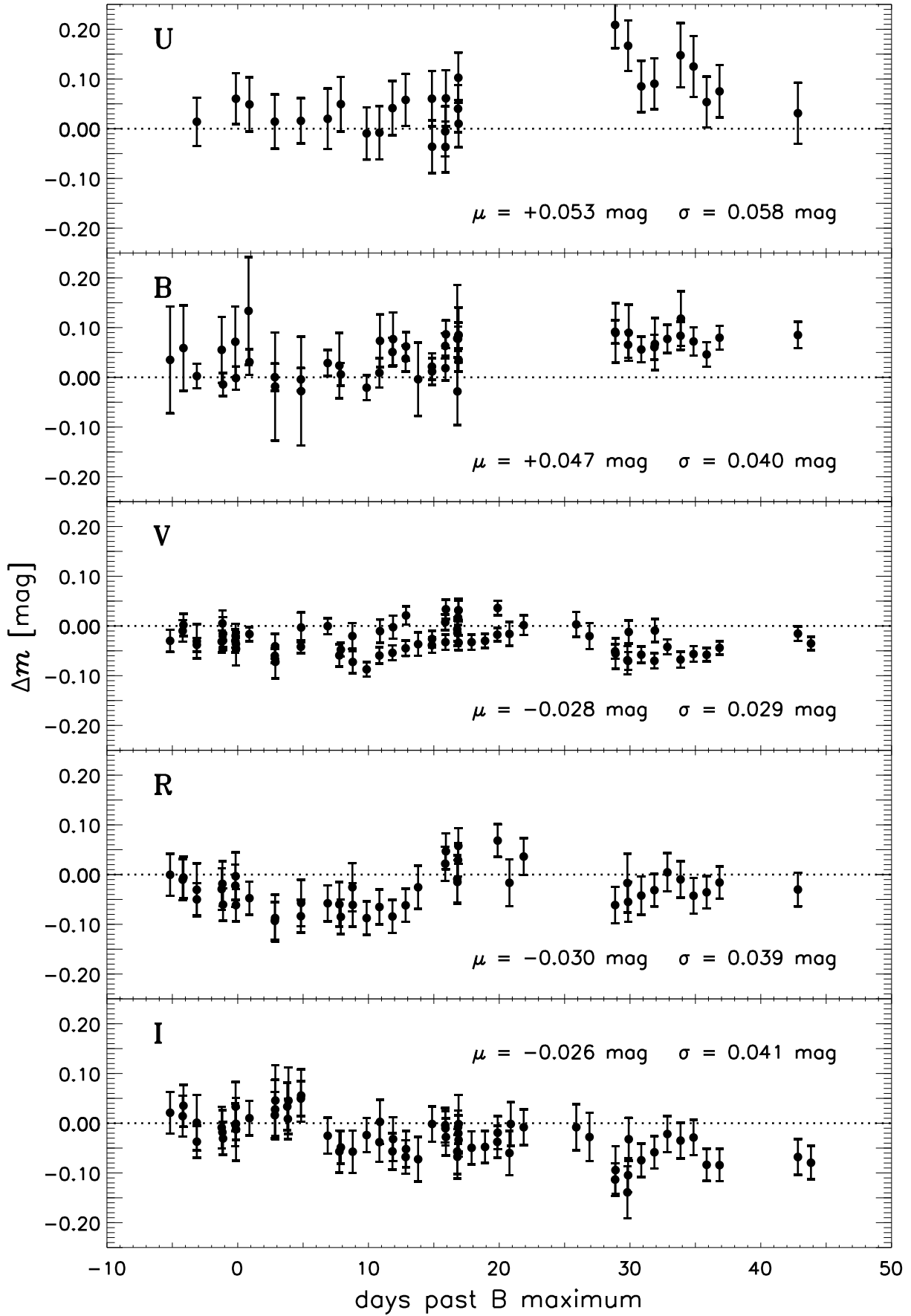}
\caption{Comparison of optical photometry with that of Suntzeff et
al. (1999). Magnitude differences (this paper - Suntzeff et al.) in
$UBVRI$ are plotted. The mean difference, $\mu$, and the dispersion,
$\sigma$, are also listed for each passband.
\label{figphotcomp}}
\end{figure*}

In comparing our photometry with the independent data set of
\markcite{sun99}Suntzeff et al. (1999), we find excellent agreement.
The times of maxima in the various passbands are quite consistent
given the stated uncertainties, as are the light curves in general.
In Figure \ref{figphotcomp}, we compare the photometry directly by
plotting the difference between our magnitudes and those of
\markcite{sun99}Suntzeff et al. (1999). To make this comparison we
have spline-interpolated the Suntzeff et al. (1999) light curves to
the times of our observations, and have only computed differences when
observations were within two days of each other to ensure the accuracy
of the interpolation. The uncertainty in the magnitude difference was
taken as the quadrature sum of the stated photometric
uncertainties. As in the case of the photometry itself, the difference
uncertainties are significantly correlated (due to the uncertainty in
the comparison stars, for instance). The largest differences occur, as
expected, in the $U$-band, and in general the agreement between the
two data sets is best near maximum light. Given the difficulties
particular to supernova photometry, the consistency in the light
curves is reassuring. Because small systematic differences in
photometry can have a magnified effect in distance determination
(through the reddening, for example), proper accounting of the
(correlated) photometric errors is vital in order to obtain consistent
results.

One important result from the photometry is that the observed color of
the supernova at maximum, $(B-V)_{\rm Bmax} = 0.34 \pm 0.04$ mag, is
significantly redder than typical SNe Ia, which have $(B-V)_{\rm Bmax}
\simeq 0.00 \pm 0.04$ mag \markcite{scha95}(Schaefer 1995). Very
underluminous supernovae such as SN 1991bg can have quite red
intrinsic colors at maximum, but they also show distinct spectroscopic
peculiarities. The absence of any such peculiarities in SN 1998bu
suggests that this red color is not intrinsic to the supernova but
rather a result of interstellar extinction along the line of
sight. Supporting evidence for this extinction is presented in Section
\ref{sectmlcs}.

\begin{figure*}
\plotone{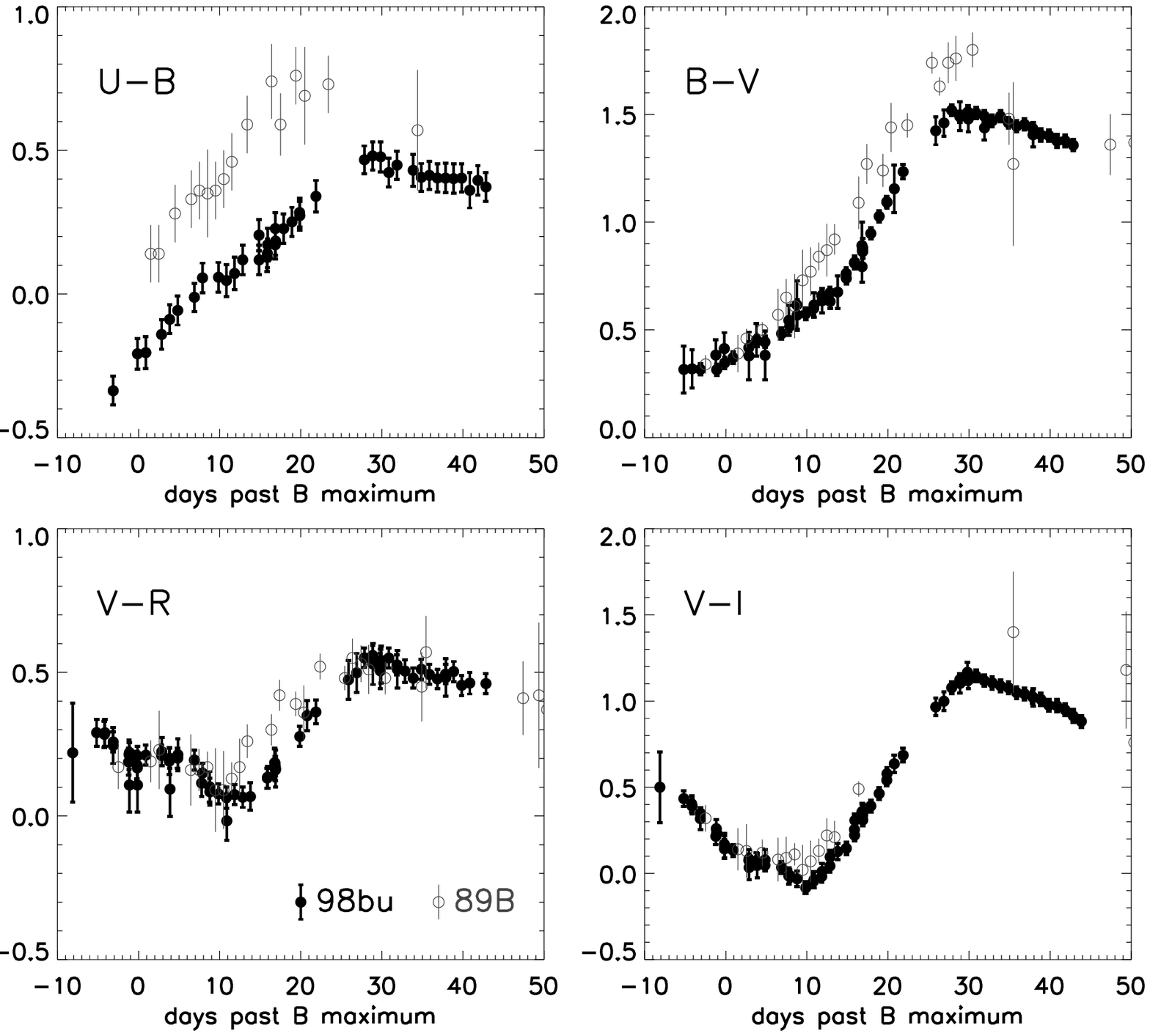}
\caption{Color curves for SN 1998bu (\emph{filled circles}) compared
with those of the reddened SN 1989B (\emph{open circles}, Wells et
al. 1994). \label{figcolcurve}}
\end{figure*}

The optical color curves of SN 1998bu are also quite typical, as shown
in Figure \ref{figcolcurve}, where we present the color evolution of
SN 1998bu compared to SN 1989B, also a spectroscopically normal SN Ia
\markcite{wel94}(Wells et al. 1994).  Both of these supernovae had a
similar $B-V$ color index at maximum light and the general shape of
the color evolution of these two supernovae are in reasonable
agreement. The slope of the $B-V$ rise is measurably different,
resulting from the fact that SN 1989B was a faster-declining object.
The other striking difference is the offset of the $U-B$ curves for
the two supernovae. This could be a result of photometric
uncertainties in calibrating the $U$ band, where detector
sensitivities and filter transmissions can differ substantially from
one site to another and require careful calibration
\markcite{sun99}(see the discussion by Suntzeff et al. 1999). However,
it may also point to interesting diversity in the $U$-band
characteristics of SNe Ia, or diversity in the selective-to-total
extinction properties of dust. A more detailed investigation of SNe Ia
light curves in the near ultraviolet is warranted. This may be
particularly important for observations of SNe Ia at high redshift
where observations at optical wavelengths probe the rest-frame
ultraviolet. Without a thorough understanding of SNe Ia $U$-band
properties, cosmological inferences based on rest-frame $U$-band light
curves are suspect.

In the infrared, the light curves of SN~1998bu match the $JHK$
templates developed by \markcite{eli85}Elias et al. (1985), as shown
in Figure \ref{figirphot}, where we have fit the templates to the data
by adjusting them independently in magnitude and together in time. The
bright second maximum typical of SNe Ia in the infrared passbands
shown in the templates is clearly observed in SN 1998bu, as is the
$J$-band deficit. The best-fit magnitude offsets to the templates are
as follows: $J = 12.14 \pm 0.09$, $H = 11.99 \pm 0.05$, and $K = 12.04
\pm 0.05$. We find that the fiducial time $t_0$ as defined by
\markcite{eli85}Elias et al. (1985) is about 3 days before maximum
light in $B$. This differs slightly from the Elias et al. result,
which suggested $t_0$ was roughly 5 days before maximum light;
however, only three supernovae were used in that determination, so it
would not be surprising if there were some variance. A larger sample
of infrared light curves, especially with observations near the first
maximum, would be useful. We note that \markcite{mei99}Meikle \&
Hernandez (1999) show a large amount of infrared photometry of SN
1998bu near optical maximum; combining these data with our light curve
(which is dominated by later points) should yield one of the best
infrared light curves of a type Ia supernova.

\subsection{Multicolor Light Curve Shape Analysis\label{sectmlcs}}

The relation between luminosity and light curve shapes for SNe Ia, as
quantified by \markcite{phi93}Phillips (1993), and subsequently
\markcite{ham96a}Hamuy et al. (1996a), led to the development of
techniques to measure distances to SNe Ia from multicolor light
curves. \markcite{ham96b}Hamuy et al. (1996b) showed how $BVI$ light
curves and templates \markcite{ham96d}(Hamuy et al. 1996d) could be
combined to derive accurate distances using a $\chi^2$ analysis. In a
similar vein, \markcite{rie96a}Riess, Press, \& Kirshner (1996a)
developed the Multicolor Light Curve Shape (MLCS) method, a
statistical technique to measure distances to SNe Ia from their $BVRI$
light curves, allowing for reddening in the host galaxy. In this
approach, the light curves of a ``training set'' of supernovae with
estimated luminosities and extinctions are used to derive template
light curves for a fiducial SN Ia, along with derived correction
templates which detail the change in the light curve shapes as a
function of luminosity and extinction. We focus on MLCS distances to
SNe Ia in this paper; \markcite{phi99}Phillips et al. (1999) present
an extension to their template-fitting technique which also
incorporates reddening, and the results of applying this method to SN
1998bu are reported by \markcite{sun99}Suntzeff et al. (1999).

The orginal MLCS training set was based on nearby SNe Ia and relative
distances measured to their host galaxies via the Tully-Fisher (TF),
the surface brightness fluctuation (SBF), or the planetary nebula
luminosity function (PNLF) methods. The only requirement was that
these methods give accurate \emph{relative} distances to the
galaxies. Once trained, the method can be used on the light curves of
a SN Ia, to determine the luminosity difference, $\Delta$, between
that supernova and the fiducial ($\Delta = 0$) supernova as well as a
derived extinction to the supernova. Application of MLCS to a sample
of more distant supernovae indicated the effectiveness of this
procedure. The dispersion in SN Ia distances about the Hubble line was
reduced from $\sigma \simeq 0.4$ mag in the standard candle assumption
to $\sigma \simeq 0.12$ mag with MLCS. The technique also demonstrated
the Hubble law was applicable to large distances corresponding to
velocities $cz \simeq 30,000 \;{\rm km \; s^{-1}}$, as was also shown
by \markcite{ham96b}Hamuy et al. (1996b).

However, uncertainties in these secondary distance determinations,
inherent difficulties in estimating the extinction to some supernovae,
and the small number of training set objects caused some problems in
the derived MLCS SN Ia distance scale, as pointed out by
\markcite{sah97}Saha et al. (1997).  To combat the major problems,
\markcite{rie98a}Riess et al. (1998a) presented a new version of MLCS
in which the relative distances for the training set objects were
derived from their host galaxy recession velocities and the Hubble law
for galaxies with redshifts $cz \geq 2500 \; {\rm km \; s^{-1}}$. In
addition, estimates of the extinction to the training set supernovae
were refined, and effects of extinction on the \emph{shape} of the
light curve based on temporal variations in the effective
selective-to-total extinction ratios from the evolving supernova
spectrum \markcite{nug99}(Nugent, Kim, \& Perlmutter 1999) were also
included. The procedure detailed in Riess et al. (1998a) was
restricted to $B$ and $V$ light curves up to 40 days past maximum, to
be applied to the high-redshift supernovae that are the focus of that
paper, but a procedurally identical version (with one exception) using
full $BVRI$ light curves is used in this paper. Here we have used a
``wide'' choice of the \emph{a priori} extinction distribution, with
$\sigma(A_V) \simeq 1$ mag rather than a distribution which overly
simplified the models of \markcite{hat98}Hatano, Branch, \& Deaton
(1998). We found that our distribution based on those models predicts
many fewer significantly reddened supernovae than are
observed. Applying a Bayesian filter based on that particular
distribution leads to underestimated extinctions in reddened
supernovae and produces biased distances. We have reverted to the less
restrictive prior distribution that was used by Riess, Press, \&
Kirshner (1996a), so that the posterior distribution is shaped
primarily by the observations rather than by the prior (cf. Figure
\ref{figav}).

The MLCS analysis fits the light curves with templates that are a
function of luminosity and extinction. To determine the peak
brightness of the supernova (in the $V$ band which is arbitrarily
chosen as the default), rather than using just the points near maximum
light, the whole light curve is used, through a weighted average of
the difference between the light curve and the best-fit template. We
designate this weighted average $\langle V_{\rm Bmax}\rangle$ to
differentiate it from the estimate of maximum light brightness based
on only the points near the time of maximum light, $V_{\rm
Bmax}$. Again, we note that the \emph{time} of maximum light is
defined in the $B$ band, such that both $V_{\rm Bmax}$ and $\langle
V_{\rm Bmax}\rangle$ describe the $V$ magnitude at the time of peak
$B$ luminosity. The difference between $V_{\rm Bmax}$ and $\langle
V_{\rm Bmax}\rangle$ is quite small in all cases, but disregarding the
distinction would make it appear as though there are discrepancies in
derived distances at the level of $\sim 0.02$ mag.

The MLCS analysis of a set of supernova light curves results in three
major parameters: $\langle V_{\rm Bmax}\rangle$, $A_V$, and
$\Delta$. Application of the MLCS method on the $BVRI$ light curves of
SN 1998bu yields a maximum light brightness $\langle V_{\rm
Bmax}\rangle = 11.89$, which is quite consistent with the result from
just the points near maximum, $V_{\rm Bmax} = 11.88 \pm 0.02$.  The
derived extinction is $A_V = 0.94$ mag and the luminosity difference
$\Delta = 0.02$ mag (i.e., the supernova was 2\% less luminous than
the fiducial). For the purposes of comparing supernovae and measuring
distances, we define $m_V \equiv \langle V_{\rm Bmax}\rangle - A_V$ as
the extinction-corrected maximum light apparent magnitude of the
supernova in the $V$ band.  We further define the quantity $m^0_V
\equiv \langle V_{\rm Bmax}\rangle - A_V - \Delta$, which would be the
maximum light apparent $V$-band brightness of the supernova had it
been free of absorption \emph{and} of fiducial luminosity. For SN
1998bu, then, we have
\begin{equation}
m_V = 10.95 \pm 0.18 \quad {\rm and}
\label{eqnmv98bu}
\end{equation}
\begin{equation}
m^0_V = 10.93 \pm 0.18,
\label{eqnm0v98bu}
\end{equation}
where the uncertainties are primarily due to the (correlated)
uncertainties in the derived luminosity correction and extinction,
with the uncertainty in the extinction ($\pm 0.15$ mag) being the
dominant component.

The derived extinction can be compared directly with the supernova's
red color at maximum light. Assuming that a typical unreddened SN Ia
has $(B-V)_{\rm Bmax} \simeq 0.00 \pm 0.04$ \markcite{scha95}(Schaefer
1995) implies a color excess for SN 1998bu of $E(B-V) = 0.34 \pm 0.06$
mag. Adopting (at maximum light) $R_V = 3.1$ yields $A_V = 1.05 \pm
0.19$ mag, fully consistent with the MLCS extinction derived from the
full $BVRI$ light curves. We note that the expected extinction from
our Galaxy along the line of sight to SN 1998bu is small, $E(B-V) =
0.025$ mag \markcite{schl98}(Schlegel, Finkbeiner, \& Davis 1998), so
that the bulk of the reddening is from M96
itself\footnote{\markcite{stan98}Stanek (1998) and
\markcite{arc99}Arce \& Goodman (1999) have recently concluded that
the Galactic reddening maps of \markcite{schl98}Schlegel, Finkbeiner,
\& Davis (1998) might overestimate the extinction in regions where
$E(B-V) \gtrsim 0.15$ mag. The Galactic extinction towards M96 is well
below this level, so this should not be a major concern.}. The $B-V$
color of SNe Ia is also generally quite uniform at $\sim 35$ days past
maximum light, with $B-V \simeq 1.1 \pm 0.1$ mag
\markcite{lir95}\markcite{rie98a}\markcite{phi99}(Lira 1995; Riess et
al. 1998a; Phillips et al. 1999). The observed color of SN 1998bu at
that time, $B-V = 1.48 \pm 0.04$ mag, also implies a color excess
consistent with the extinction derived from the full MLCS
analysis. \markcite{sun99}Suntzeff et al. (1999) derive a total
reddening for SN 1998bu of $E(B-V) = 0.37$ mag based on the $B-V$ and
$V-I$ color evolution \markcite{phi99}(Phillips et al. 1999), which is
consistent with our results.

Further evidence comes from the infrared light curves, where effects
of dust are expected to be small. By comparing our IR observations
with those of other well-observed type Ia events and assuming that the
optical-IR colors are constant for normal SN Ia, we can derive another
check on the inferred extinction. For example, the $V$ maximum of SN
1981B was fainter by 0.08 mag than the $V$ maximum of SN 1998bu, while
in $K$-band the SN 1981B light curve was fainter than the SN 1998bu
light curve by 0.87 mag \markcite{eli81}(Elias et al. 1981). Using the
extinction law of \markcite{car89}Cardelli, Clayton, \& Mathis (1989)
with $R_V = 3.1$, the difference in visual extinction between the two
supernovae is given by $\Delta A_V = (\Delta K-\Delta
V)/0.886$. Hence, the relative extinction between 1981B and 1998bu is
$\Delta A_V = 0.89$ mag. Unfortunately, there are few type~Ia
supernovae with low extinctions and good IR light curves which can be
used in this way. Combining infrared and optical data of SN~1980N and
SN~1981D, both in NGC~1316 \markcite{ham91}(Hamuy et al. 1991), gives
a relative extinction of 0.81 mag and a rough lower limit consistent
with our derived value. SN~1989B was highly extinguished as is
SN~1998bu. The visual magnitude difference between these two is 0.11
mag (SN 1998bu is brighter) and the $K$ difference is $-0.07$ mag (SN
1998bu is fainter), suggesting that there is 0.20 mag less visual
extinction to SN~1998bu than SN~1989B. \markcite{wel94}Wells et
al. (1994) found a color excess to SN~1989B of $E(B-V) = 0.37$ mag,
meaning the extinction to SN~1998bu would be $A_V = 0.95$ mag. All of
the estimates from the infrared photometry are consistent with a total
visual extinction to SN~1998bu of $A_V = 0.9 \pm 0.2$ mag, supporting
the value derived in the MLCS analysis.

These checks based on the color excesses at a number of wavelengths
from $B$-band to $K$-band are consistent with the view that the shape
of the extinction curve is likely close to the standard Galactic law
\markcite{rie96b}(Riess, Press, \& Kirshner 1996b), though the
absolute normalization is not constrained. Spectrophotometric
observations of SNe Ia have been used to determine the normalization,
with $R_V \simeq 3.1$ at maximum light, as well as the temporal
variation of the effective $R_V$ due to the evolution of the supernova
spectrum \markcite{nug99}(Nugent, Kim, \& Perlmutter 1999). The
$U$-band photometry still provides some cause for concern; if the blue
$U-B$ color is due to differences in the properties of the absorbing
dust, the estimated extinction may be incorrect.

Independent checks of the MLCS-derived extinction are valuable. One
such check is the presence of interstellar \ion{Na}{1} D and
\ion{Ca}{2} H \& K absorption in high-resolution spectra, which was
reported by \markcite{mun98}Munari et al. (1998) and
\markcite{cen98}Centurion et al. (1998). The equivalent width of the
\ion{Na}{1} D1 ($\lambda 589.0$ nm) absorption lines reported by
\markcite{mun98}Munari et al. (1998) were 0.019 nm and 0.035 nm at
velocities corresponding to our Galaxy and M96, respectively, and we
confirm these measurements even with our low-dispersion spectra. Using
the \markcite{mun97}Munari \& Zwitter (1997) calibration of the
correlation between the equivalent width and reddening they derive
color excesses of 0.06 and 0.15 mag, for a total reddening of $E(B-V)
= 0.21$ mag. However, the relation between the absorption-line
equivalent widths and the reddening has a large scatter, with a
typical dispersion of 0.15 mag in $E(B-V)$ for multi-component lines
\markcite{mun97}(Munari \& Zwitter 1997), so that these values do not
contradict the extinction inferred from the MLCS analysis
(cf. discussion by \markcite{sun99}Suntzeff et al. 1999).

We can also check our derived luminosity with other techniques. The
relation between light curve shape and luminosity was pioneered by
\markcite{phi93}Phillips (1993) and
\markcite{ham95}\markcite{ham96a}Hamuy et al. (1995, 1996a) using the
quantity $\Delta m_{15}(B)$, which parameterizes the $BVI$ light
curves in terms of the $B$ magnitude decline of the supernova over the
fifteen days after maximum light. From our light curve, we measure
$\Delta m_{15}(B) = 1.02 \pm 0.04$, which agrees very with
\markcite{sun99}Suntzeff et al. (1999), who found $\Delta m_{15}(B) =
1.01 \pm 0.05$. Direct comparison of the luminosity correction is made
difficult because of effects of extinction on the light curve shape,
as well as differences in our respective fiducial
templates. Nevertheless, we can measure $\Delta m_{15}(B)$ from the
MLCS fiducial template ($\Delta = 0$, $A_V = 0$) which yields $\Delta
m_{15}(B) = 1.08$. Thus the measured values of $\Delta m_{15}$ and
$\Delta$, both of which imply SN 1998bu to be quite close to the MLCS
fiducial template, indicate consistency in the two approaches. This is
not surprising, since both methods use the shape of the observed light
curve in a similar fashion.

An alternative approach was described by \markcite{nug95}Nugent et
al. (1995), who presented correlations between spectral features and
intrinsic SNe Ia luminosity. In particular they define two indicators:
$\mathcal{R}($\ion{Si}{2}$)$, the ratio of the depths of \ion{Si}{2}
absorption lines observed at 580 and 615 nm; and
$\mathcal{R}($\ion{Ca}{2}$)$, the flux ratio of the continuum levels
just blue and red of the \ion{Ca}{2} H \& K absorption. Our
maximum-light spectra of SN 1998bu yield $\mathcal{R}($\ion{Si}{2}$) =
0.23 \pm 0.02$ and $\mathcal{R}($\ion{Ca}{2}$) = 1.47 \pm
0.04$. \markcite{rie98b}Riess et al. (1998b) present linear relations
between $\Delta$ and both $\mathcal{R}($\ion{Si}{2}$)$ and
$\mathcal{R}($\ion{Ca}{2}$)$, which yield a mean luminosity correction
for SN 1998bu of $\Delta = -0.02 \pm 0.16$ and $\Delta = 0.06 \pm
0.22$, respectively. These agree well with the luminosity correction
result derived in the MLCS analysis.

All indications thus suggest that SN 1998bu was an intrinsically
normal type Ia supernova, significantly extinguished by dust along the
line of sight. With the MLCS analysis we determine the peak brightness
($\langle V_{\rm Bmax} \rangle = 11.89$), the luminosity
correction ($\Delta = 0.02$ mag), and the extinction ($A_V = 0.94$
mag), so that SN 1998bu can be used to calibrate the SN Ia distance
scale.

\section{The Distance Scale}

\subsection{Hubble-Flow SNe Ia}

Constructing the Hubble diagram requires a sample of well-observed SNe
Ia in the Hubble flow where errors due to peculiar velocities are
expected to be small, and which is analyzed in exactly the same way as
the local calibrators. Our MLCS sample consists of 42 SNe Ia, 26 from
the Cal\'an/Tololo supernova search
\markcite{ham93}\markcite{ham96c}(Hamuy et al. 1993, 1996c) and 16
from the CfA supernova monitoring campaign \markcite{rie99}(Riess et
al. 1999). The only further selection criteria we have imposed (other
than those inherent in the two data sets) is a cut in the host-galaxy
recession velocity, which has been corrected to the cosmic microwave
background (CMB) frame\footnote{Heliocentric redshifts for the host
galaxies were first transformed to the Local Group rest frame by
adding (-30, 297, -27) ${\rm km \; s^{-1}}$ in Galactic Cartesian
coordinates \markcite{dev91}\markcite{lyn88}(de Vaucouleurs et
al. 1991; Lynden-Bell \& Lahav 1988). The recession velocities in the
Local Group frame were then transformed to the CMB rest frame by
adding (10, -542, 300) ${\rm km \; s^{-1}}$ \markcite{smo92}(Smoot et
al. 1992).}. We have excluded supernovae in galaxies with $cz < 2500
\; {\rm km \; s^{-1}}$ where peculiar motions become increasingly
important. Additionally, we have excluded supernovae in galaxies with
$\log cz \; [{\rm km \; s^{-1}}] > 4.5$, where the relation between
luminosity-distance and redshift begins to be non-linear at a level
which could affect our results \markcite{sch98}(Schmidt et al. 1998).

The Hubble-flow sample consists of the following SNe Ia:
SN 1990O, SN 1990af, SN 1991U, SN 1991ag, SN 1992J, SN 1992K, SN
1992P, SN 1992ae, SN 1992ag, SN 1992al, SN 1992aq, SN 1992au, SN
1992bc, SN 1992bg, SN 1992bh, SN 1992bk, SN 1992bl, SN 1992bo, SN
1992bp, SN 1992br, SN 1992bs, SN 1993B, SN 1993H, SN 1993O, SN 1993ac,
SN 1993ae, SN 1993ag, SN 1993ah, SN 1994M, SN 1994Q, SN 1994S, SN
1994T, SN 1995D, SN 1995E, SN 1995ac, SN 1995ak, SN 1995bd, SN 1996C,
SN 1996Z, SN 1996bl, SN 1996bo, and SN 1996bv. We emphasize that all
of these supernovae have well-sampled multicolor CCD light curves, with
photometry obtained, reduced, and transformed to the standard system in
a similar fashion, an essential feature which allows us to combine the
data sets.

We use MLCS to turn these SNe Ia into standard candles, so that 
the extinction- and luminosity-corrected maximum light absolute
magnitude, $M^0_V = m^0_V - \mu$, is a constant, where $\mu$ is the
distance modulus. Using the definition of the distance modulus,
\begin{equation}
m^0_V - M^0_V = \mu = 5 \log d + 25 = 5 \log \frac{cz}{H_0} + 25, 
\end{equation}
where $d$ is the distance in Mpc, $cz$ is the recession
velocity in $\rm{km \; s^{-1}}$, and $H_0$ is the Hubble constant
measured in its conventional units of $\rm{km \; s^{-1} \; Mpc^{-1}}$,
we get the result
\begin{equation}
\log cz - 0.2 m^0_V = \log H_0 - 0.2 M^0_V - 5.
\end{equation}
Since the absolute magnitude of the fiducial SN Ia is taken to be
constant, we can determine that
\begin{equation}
\log cz = 0.2 m^0_V + a_V.
\end{equation}
Here $a_V \equiv \log H_0 - 0.2 M^0_V - 5$ is ``the intercept of the
ridge line'' and is a constant which can be determined from
observations of Hubble-flow SNe Ia alone.

\begin{figure*}
\plotone{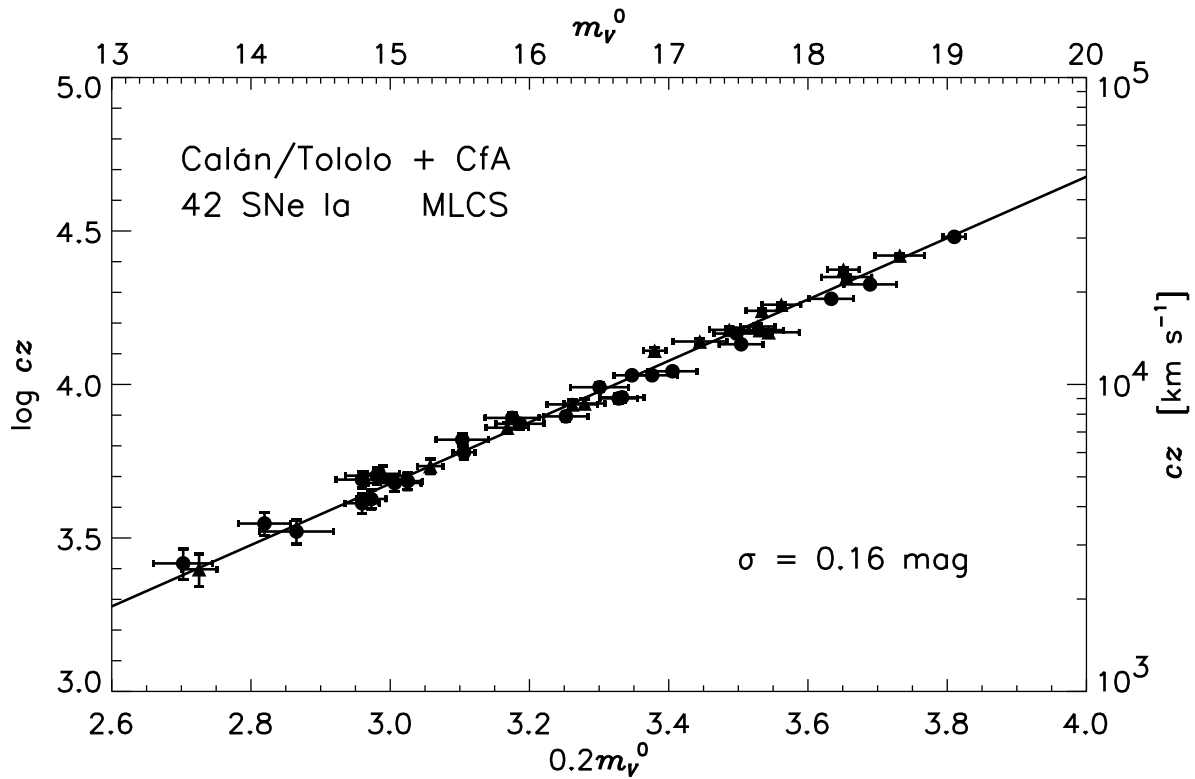}
\caption{Apparent magnitude-redshift relation for 42 Hubble-flow SNe
Ia which were corrected for extinction and to a fiducial luminosity
with a multicolor light curve shape (MLCS) analysis. The supernovae
are from the Cal\'an/Tololo (Hamuy et al. 1996c) and CfA (Riess et
al. 1999) data sets. Supernovae in late-type ($\geq$Sa) galaxies are
shown with circles, those in early-type (E/S0) galaxies are shown with
triangles. The best-fit ridge line is shown, $\log cz = 0.2 m^0_V +
0.6772 (\pm 0.0049)$. The dispersion about the best-fit line is
$\sigma = 0.16$~mag. \label{fighubflow}}
\end{figure*}

\begin{figure*}
\plotone{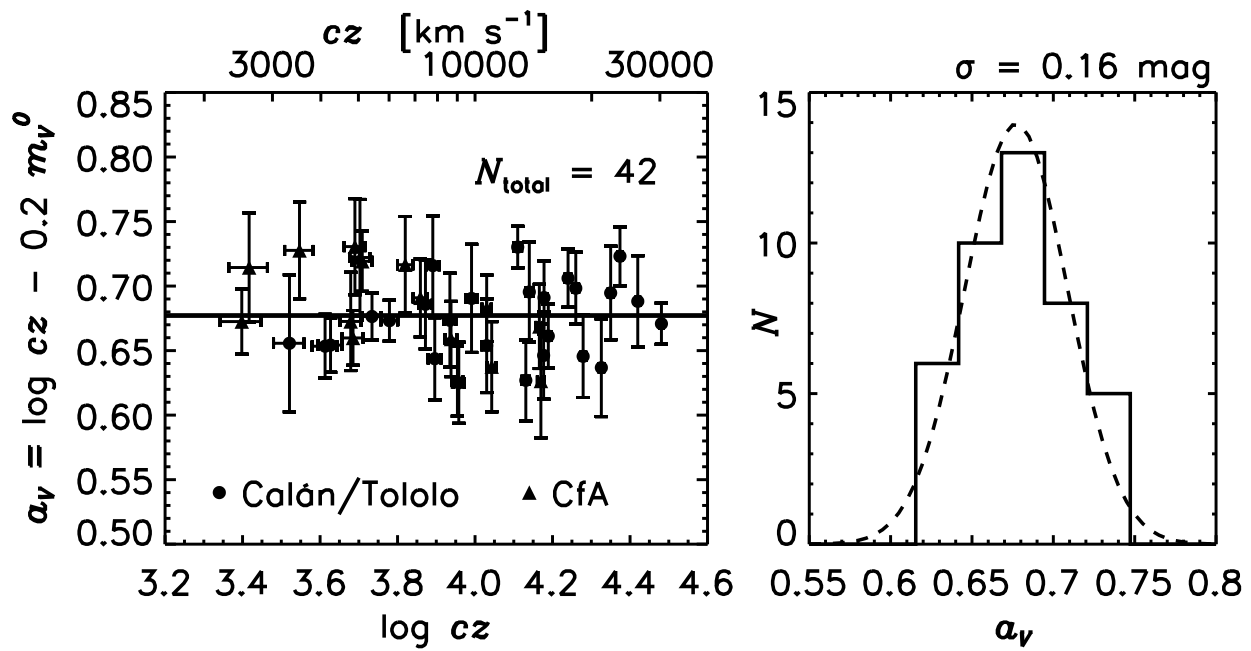}
\caption{Ridge-line intercept versus recession velocity for the
Hubble-flow sample (\emph{left}) and ridge-line intercept histogram
(\emph{right}). The best-fit mean intercept and its formal uncertainty
are $a_V = 0.6772 \pm 0.0049$. \label{fighf}}
\end{figure*}

In Figure \ref{fighubflow} we present this relation for our sample of
42 Hubble-flow SNe. The small scatter allows us to solve precisely for
the intercept, $a_V$, as shown more clearly in Figure \ref{fighf}. The
best-fit intercept (which is simply the mean of $\log cz - 0.2 m^0_V$)
using all the supernovae is $a_V = 0.6772 \pm 0.0049$, where the
uncertainty is the formal standard error in the mean, and assumes the
residuals from each supernova are normally distributed and
independent. We assign a $\pm 300 \; {\rm km \; s^{-1}}$ 1$\sigma$
uncertainty to the redshift to account for the contribution of
peculiar motions \markcite{rie96a}(Riess, Press, \& Kirshner
1996a). The dispersion about the mean is $\sigma(a_V) = 0.0317$; this
corresponds to a dispersion in magnitudes (obtained simply by
multiplying by 5) of $\sigma = 0.16$ mag, implying only an 8\%
relative distance uncertainty per object\footnote{Some of this
uncertainty arises from the uncertainty in the redshift due to
peculiar velocities. Our sample has an error-weighted mean redshift of
roughly 10,000 ${\rm km \; s^{-1}}$ so that the adopted 300 ${\rm km
\; s^{-1}}$ uncertainty corresponds to a 3\% distance uncertainty on
average. This means the actual relative distance uncertainty intrinsic
to the MLCS analysis of the supernovae is only 7\% per object.}. Our
derived intercept depends upon the choice for the fiducial luminosity
($\Delta = 0$) supernova. As long as the comparison with the local
calibrators is made with the same choice, there is no
problem. However, when comparing this set of Hubble-flow SNe with
those analyzed by a different technique, such as the $\Delta
m_{15}(B)$ method \markcite{sun99}\markcite{phi99}(Suntzeff et
al. 1999; Phillips et al. 1999), there will likely be an offset in
$a_V$ due to the different choices of a fiducial luminosity. Such an
offset will also be reflected as the same offset for the fiducial
absolute magnitude, $M^0_V$.

The quoted statistical uncertainty in the measurement of $a_V$ is
quite small, and it is surely underestimated. We assumed that each
supernova distance is independent, whereas in reality there exists
some covariance. Thus, the true uncertainty in the mean does not
simply decrease as $\sigma/\sqrt{N}$, but rather levels off due to a
floor caused by systematic uncertainties. It is thus important to
estimate at what level this floor is reached. The formal uncertainty
in $a_V$ corresponds to $\pm 0.025$ mag or just over one percent
uncertainty in the Hubble constant arising just from the Hubble-flow
supernovae.

Sample differences between the Hubble-flow and calibrating SNe Ia are
one potential source of systematic uncertainty at this level. The
present sample is necessarily imperfect; one difference occurs due to
the host-galaxy type: our Hubble-flow sample includes supernovae in
both early-type and late-type hosts, while hosts of the local
calibrators have Cepheid distances, and thus are of late-type
only. \markcite{ham95}Hamuy et al. (1995) have shown that SNe Ia in
E/S0 galaxies are systematically fainter than those in spirals or
irregulars (we refer to galaxies classified Sa or later, including
irregulars, as ``spirals'' in what follows). More accurately, the
highest luminosity SNe Ia are found only in spirals
\markcite{rie99}(Riess et al. 1999), perhaps implying a relation
between recent star formation and the brightest SNe Ia. Nevertheless,
SNe Ia brightness in both early-type and late-type galaxies correlate
similarly with light curve shape, so that an MLCS (or $\Delta
m_{15}(B)$) analysis will still correct SNe Ia in spirals and
ellipticals to the same fiducial luminosity without using any
information on the galaxy morphology, as demonstrated by
\markcite{sch98}Schmidt et al. (1998). To further test this we have
divided our sample into two subsets, early-type and late-type hosts,
solving independently for the intercept. The results are presented in
Table \ref{tabav}. We see that after application of MLCS the offset
between $a_V$ in early-type and late-type galaxies is inconsistent
with zero at only the 1.5$\sigma$ level. Determining whether this
difference is significant will require a larger sample. However, this
offset would lead to a difference in the derived Hubble constant (for a
fixed $M^0_V$) of $\sim 3\%$ (in the sense that early-type galaxies
yield the slightly higher value).

\begin{deluxetable}{cccc}
\tablecaption{Intercepts of the Ridge Line for Hubble-Flow SNe Ia. 
\label{tabav}}
\tablehead{\colhead{Sample} & \colhead{$N$} & \colhead{$a_V$} & 
\colhead{$\sigma$~[mag]} }
\startdata
All & 					42 & $0.6772 \pm 0.0049$ & 0.16 \nl
\nl
Late-type ($\geq$Sa) &			25 & $0.6716 \pm 0.0067$ & 0.17 \nl
Early-type (E/S0) &			17 & $0.6854 \pm 0.0067$ & 0.14 \nl
\nl
$cz \geq 7000 \; {\rm km \; s^{-1}}$ &	28 & $0.6712 \pm 0.0059$ & 0.16 \nl
$cz < 7000 \; {\rm km \; s^{-1}}$ &	14 & $0.6892 \pm 0.0081$ & 0.15 \nl
\enddata
\end{deluxetable}

Another potential source of systematic uncertainty in the measurement
of $a_V$ is the effect of galaxy peculiar velocities and flows. We
have transformed the measured host-galaxy recession velocity to the
frame at rest with respect to the CMB frame, but distortions of the
velocity field will result in errors in our derived intercept. Some
peculiar velocity studies \markcite{gio98}(e.g., Giovanelli et
al. 1998) have indicated convergence of the flow field relative to the
Local Group with the CMB dipole at redshifts $cz \simeq 4000 \; {\rm
km \; s^{-1}}$, though others do not \markcite{lau94}(e.g., Lauer \&
Postman 1994). Our full sample is cut at $cz \geq 2500 \; {\rm km \;
s^{-1}}$, so flows may be important at the lowest redshifts. To check
this we examined a subset of data with $cz \geq 7000 \; {\rm km \;
s^{-1}}$ where such motions should have a much smaller effect. As
Table \ref{tabav} shows, there is no signifcant difference in $a_V$
between our full sample and the sample restricted to $cz \geq 7000 \;
{\rm km \; s^{-1}}$.

There is a larger difference in comparing the more distant sample with
the remaining SNe Ia, i.e. those with $cz < 7000 \; {\rm km \;
s^{-1}}$. In this case the offset in $a_V$ differs from zero at
$1.8\sigma$. \markcite{zeh98}Zehavi et al. (1998) have interpreted
this result as the effect of a local void, whereby we live in a
slightly underdense region compared to the average density of the
universe, leading to a larger nearby expansion rate. For a fixed
$M^0_V$ the difference in $a_V$ would lead to a ``local'' Hubble
constant larger than the global value by $\sim 4\%$. Again, more
Hubble-flow SN Ia light curves will help determine whether this offset
is real, a statistical fluke, or an artifact of the analysis
technique. Since the large majority of our sample (28 of 42) have $cz
\geq 7000 \; {\rm km \; s^{-1}}$, the mean is more reflective of the
global value.

We have performed an additional test of our derived intercept by
employing a simple geometric flow model which includes the effects of
nearby mass concentrations such as the Virgo Cluster, the Great
Attractor, and the Shapley Supercluster on the velocities of the
supernovae host galaxies \markcite{mou99}(Mould et al. 1999). Using
the position and redshifts of our galaxy sample, this model predicts
that we underestimate the Hubble constant by $\sim 2\%$ in assuming
that the Hubble-flow galaxies are at rest with respect to the CMB
frame. 

Additional sources of uncertainty may remain, for instance due to
correlations in the MLCS analysis arising from the training set and
construction of the templates, or uncertainties in the calibration of
the photometric system. A larger sample of Hubble-flow SNe Ia would be
helpful to determine what unidentified systematics may remain and
at what level they affect our conclusions.

Given that the identified systematic errors in the Hubble-flow SNe
do not yet definitively suggest a bias in our derived intercept, we
use these results only as a guide to the size of the systematic
uncertainty. Based on these explorations, a reasonable estimate of the
1$\sigma$ systematic uncertainty in the Hubble-flow supernovae would
be $\pm3\%$ in the Hubble constant. Our best estimate for the
intercept of the ridge-line and its total uncertainty is then $a_V =
0.6772 \pm 0.0120$. As we discuss below, this uncertainty in the
Hubble-flow SNe Ia is dwarfed by both statistical and systematic
uncertainties in the Cepheid-calibrated supernovae and our measurement
of $M^0_V$ \markcite{ham96b}\markcite{rie96a}(Hamuy et al. 1996b;
Riess, Press, \& Kirshner 1996a).

\subsection{Cepheid-Calibrated SNe Ia\label{sectcephcal}}

To measure the Hubble constant we need both the intercept of the ridge
line, $a_V$, and the maximum light absolute magnitude of our fiducial
SN Ia, $M^0_V$. We use a sample of SNe Ia in galaxies with distances
measured via Cepheids. SN 1998bu in NGC 3368 is only the most recent
example, others are listed by \markcite{sah97}Saha et al. (1997) in
their Table 6: SN 1895B and SN 1972E in NGC 5253, SN 1937C in IC 4182,
SN 1960F in NGC 4496A, SN 1981B in NGC 4536, and SN 1990N in NGC
4639. Their table also includes SN 1989B in NGC 3627, but the distance
to this galaxy is only inferred from distances to other galaxies in
the Leo group (including NGC 3368). We restrict our sample to the best
cases: supernovae in galaxies whose distances are directly measured by
Cepheids rather than distances to groups or
clusters. \markcite{scha98}Schaefer (1998) has also recalibrated the
light curve of SN 1974G in NGC 4414, for which the host-galaxy Cepheid
distance has been measured by the \emph{HST} Distance Scale Key
Project \markcite{tur98}(Turner et al. 1998).

The supernovae we use as calibrators are only those which are measured
and analyzed in the same way as our Hubble-flow sample, to avoid
systematic errors. For this reason, we only consider SNe Ia which have
photoelectrically measured multicolor light curves. As our
observations of SN 1998bu have shown, even CCD data measured and
calibrated similarly can yield discrepant photometry depending on the
details of the telescope, detector, and filters. Such calibration
problems, as well as problems of galaxy background subtraction and
transformation to the Landolt system, make photographic photometry of
supernovae subject to systematic differences in the peak brightness,
colors, and the light curve shape. The last point is crucial; we must
be certain that the observed light curve shape is an intrinsic
property of the supernova, and not an artifact of systematic errors
resulting from photographic photometry
\markcite{boi91}\markcite{pie95}(Boisseau \& Wheeler 1991; Pierce \&
Jacoby 1995). While heroic efforts \markcite{scha98}(e.g., Schaefer
1998 and references therein) have been made in compiling and
reanalyzing older photographic (and even visual) light curves, the
best path to the Hubble constant lies along another route: precise
distances to well-observed objects. The drawback to a high standard
for the data is that our calibrating sample is small, consisting of
four SNe Ia: SN 1998bu, SN 1990N, SN 1981B and SN 1972E. We pay in
random error what we avoid in systematic bias and we believe this to
be a good bargain.

The Cepheid distances to the host galaxies of these four supernovae
have been measured by two \emph{HST} programs, but the general
approach among all the \emph{HST} Cepheid programs is the same. From
the derived mean magnitudes of the Cepheids and a $PL$ relation
\markcite{mad91}(Madore \& Freedman 1991), distance moduli (relative
to the LMC) can be determined. The two-color photometry allows for an
estimate of the extinction, either on a Cepheid-by-Cepheid basis or in
the mean, yielding an extinction-corrected distance modulus for the
host galaxy.

It is important to use distance moduli for the host galaxies which are
derived consistently (e.g., always using the same $PL$ relation, the
same LMC distance modulus, the same extinction prescription, etc.) so
that each supernova is on an equal footing. Thus we have not simply
used the final distance modulus quoted in the Cepheid papers, but
rather have tried to extract as uniform a set of distance moduli as we
can with limited information. This approach also allows us to more
easily consider systematic effects in the Cepheid distances. For
instance, in this section we use distance moduli with the LMC distance
fixed at $\mu_{\rm LMC} \equiv 18.50$ mag (Hereafter, all distance
moduli will have implied units of magnitudes). We \emph{do not} yet
include the uncertainty in this value because that uncertainty is
implicit in each host galaxy distance, and moreover it is perfectly
correlated, such that the derived mean absolute magnitudes will suffer
the same uncertainty. We postpone discussion and quantification of
such systematic (``external'') uncertainties to the next section.

\subsubsection{SN 1998bu in M96 (NGC 3368)}

For SN 1998bu, \markcite{tan95}Tanvir et al. (1995) discovered seven
Cepheids with well-determined light curves in M96. They derived an
extinction-corrected distance modulus of $\mu = 30.32 \pm 0.12$ with
the uncertainty coming from the photometric errors and the uncertainty
(in the mean) of the fit to the $PL$ relation. However, their
photometry was not corrected for the WFPC2 ``long/short'' exposure
effect \markcite{hil98}(Hill et al. 1998) which leads to $V$ and $I$
magnitudes systematically too bright by 0.05 mag in long exposures
such as those for Cepheid programs. We have corrected the distance
modulus of M96 for this effect, yielding $\mu = 30.37 \pm 0.12$, with
the quoted uncertainty being only the ``internal'' error. Combined
with the extinction-corrected maximum light apparent magnitude from
Equation \ref{eqnmv98bu}, we derive an extinction-corrected maximum
light absolute magnitude for SN 1998bu, $M_V = -19.42 \pm 0.22$. To
determine the absolute magnitude of our fiducial SN Ia, $M^0_V$, we
also include the derived luminosity difference $\Delta = 0.02$ mag for
SN 1998bu, which leads to $M^0_V = -19.44 \pm 0.22$.

Our absolute magnitude for SN 1998bu is fainter than that reported by
\markcite{sun99}Suntzeff et al. (1999) who found $M_V = -19.63 \pm
0.19$. This offset arises from different estimates of the
extinction. We have employed $A_V = 0.94$ mag derived from the MLCS
analysis of the $BVRI$ light curves, whereas \markcite{sun99}Suntzeff
et al. (1999) derive a total $E(B-V) = 0.37$ mag, yielding $A_V =
1.15$ mag.  This difference is, for better or worse, within the
uncertainties, and may arise partly from an offset in the intrinsic
colors in the light curve fitting methods, but some more careful
comparisons between the two methods may be necessary. Nevertheless, it
turns out that the derived Hubble constant is not very different in
the two methods.

\subsubsection{SN 1990N in NGC 4639}

For SN 1990N we have performed an MLCS analysis on $BVRI$ light curves
from \markcite{lir98}Lira et al. (1998), with the results presented in
Table \ref{tabcalsn}. We have used the extinction corrected Cepheid
distance modulus to NGC 4639 determined by \markcite{sah97}Saha et
al. (1997), $\mu = 32.03 \pm 0.22$, based on 15 high-quality
Cepheids. From this we derive an absolute magnitude for SN 1990N of
$M_V = -19.78 \pm 0.33$ and an estimate of the fiducial absolute
magnitude $M^0_V = -19.46 \pm 0.33$.

\subsubsection{SN 1981B in NGC 4536}

Our MLCS analysis of SN 1981B was based on the $BVR$ light curves of
\markcite{but83}Buta \& Turner (1983). \markcite{sah96}Saha et
al. (1996) found a total of 73 Cepheids in NGC 4536 and determined an
extinction corrected distance modulus, $\mu = 31.10 \pm 0.13$. This
leads to an absolute magnitude for SN 1981B of $M_V = -19.46 \pm 0.23$
and $M^0_V = -19.12 \pm 0.23$.

\subsubsection{SN 1972E in NGC 5253}

Finally, for SN 1972E, we have $BVI$ light curves from
\markcite{ard73}Ardeberg \& de Groot (1973) and
\markcite{lei91b}Leibundgut et al. (1991b). \markcite{sah95}Saha et
al. (1995) have presented their final analysis of Cepheids in NGC
5253, with their result $\mu = 28.08 \pm 0.10$. However, some cautions
are in order. These observations were made with the original WFPC
instrument (with spherical aberration) rather than WFPC2; also, the
$I$-band light curves were transformed from the \emph{HST} F785LP
filter rather than F814W. These differences may possibly lead to a
small systematic difference in the derived distance modulus compared
to the other host galaxies. Furthermore, the Cepheid sample is small,
as there are only 5 Cepheids with high-confidence mean magnitudes in
both $V$ and $I$. Using the derived apparent moduli in $V$ and $I$,
and estimating the Cepheid extinction in the same manner as for the
other three host galaxies, we have rederived the same distance modulus
as \markcite{sah95}Saha et al. (1995) but with a significantly larger
uncertainty, $\mu = 28.08 \pm 0.26$. Our MLCS analysis of SN 1972E
then leads to $M_V = -19.80 \pm 0.29$ and $M^0_V = -19.42 \pm 0.29$.

We present the MLCS results, Cepheid distances, and absolute
magnitudes for SN 1998bu and the three other calibrating SNe Ia in
Table \ref{tabcalsn}. We note that the estimates for $M^0_V$ are
consistent given their uncertainties, though SN 1981B seems to give a
measurably fainter value. Since the estimates are mutually consistent
and there is no \emph{a priori} reason to distrust any of them, we
take the data at face value. The error-weighted mean gives the maximum
light absolute magnitude of our fiducial SN Ia, $M^0_V = -19.34 \pm
0.13$. As with the ridge-line intercept, here again we must be wary of
the uncertainty estimate since it ignores the covariance in the
Cepheid distances. \markcite{koc97}Kochanek (1997) has shown there is
significant statistical covariance in the Cepheid distance moduli,
even beyond the common zero-point set by the LMC distance, which
arises from a number of sources including the Cepheid photometry, the
determination of mean magnitudes, and the fit to the $PL$ relation. If
we include an estimate of this statistical covariance to determine how
to combine the Cepheid distances (usually incorrectly assumed to be
independent) by using the Pearson correlation coefficient, $r \simeq
0.5$ (C. Kochanek, personal communication), our best estimate of the
fiducial absolute magnitude and its statistical uncertainty becomes
$M^0_V = -19.34 \pm 0.17$. This estimate still does not incorporate
some sources of systematic uncertainty, including the LMC distance,
which we dicuss in detail in Section \ref{secextunc}.

\begin{deluxetable}{llccccccc}
\footnotesize
\tablecaption{Cepheid-Calibrated SNe Ia \label{tabcalsn}}
\tablehead{\colhead{SN Ia} & \colhead{Galaxy} & \colhead{$\langle V_{\rm Bmax} \rangle$} & 
\colhead{$A_V$} & \colhead{$\Delta$} & \colhead{$\sigma_{\rm MLCS}$} &
\colhead{$\mu_{\rm Cepheid}$} & \colhead{$M_V$} & \colhead{$M^0_V$} }
\startdata
1998bu	& NGC 3368 & 11.89 & 0.94 & $+0.02$ & $\pm0.18$ & $30.37 \pm 0.12$
	& $-19.42 \pm 0.22$ & $-19.44 \pm 0.22$ \nl

1990N  	& NGC 4639 & 12.68 & 0.43 & $-0.32$ & $\pm0.25$ & $32.03 \pm 0.22$
	& $-19.78 \pm 0.33$ & $-19.46 \pm 0.33$ \nl

1981B	& NGC 4536 & 11.99 & 0.35 & $-0.34$ & $\pm0.18$ & $31.10 \pm 0.13$
	& $-19.46 \pm 0.23$ & $-19.12 \pm 0.23$ \nl

1972E	& NGC 5253 & \phn8.43 & 0.15 & $-0.38$ & $\pm0.13$ & $28.08 \pm 0.26$
	& $-19.80 \pm 0.29$ & $-19.42 \pm 0.29$ \nl

\cline{1-9}

Mean	& \nodata & \nodata & \nodata & \nodata & \nodata & \nodata
	& \nodata & $-19.34 \pm 0.17$ \nl
\enddata
\end{deluxetable}

\subsection{The Hubble Constant}

With estimates of the ridge-line intercept and the maximum-light
absolute magnitude of our fiducial SN Ia, we derive the Hubble
constant, from
\begin{equation}
\log H_0 = 0.2M^0_V + 5 + a_V.
\end{equation}
The mean of the four calibrating SNe Ia gives $M^0_V = -19.34
\pm 0.17$, and using all 42 Hubble-flow SNe Ia gives $a_V = 0.6772 \pm
0.0120$, which results in our best estimate of the Hubble constant,
\begin{equation}
H_0 = 64.4^{+5.6}_{-5.1} \; {\rm km \; s^{-1} \; Mpc^{-1}},
\end{equation}
where the uncertainty \emph{does not} include systematic uncertainties
in the Cepheid distance scale to be discussed below. Even so, it is
important to note that the total uncertainty in $a_V (\pm 0.0120)$ is
about three times smaller than the statistical uncertainty in $0.2
M^0_V (\pm 0.034)$. The statistical error in this small sample of
calibrating SNe Ia, arising from both the uncertainty in the
luminosity- and extinction-corrected supernova brightness and the
uncertainty in the Cepheid distance moduli, dominates the statistical
uncertainty in $H_0$ \markcite{ham96b}\markcite{rie96a}(Hamuy et
al. 1996b; Riess, Press, \& Kirshner 1996a). Reducing this statistical
error can best be accomplished by observations of additional nearby SNe
Ia and Cepheids in their host galaxies. SN 1998bu is the first example
where a new supernova has been studied in a galaxy where the Cepheid
work is already in the literature, but more will follow in the years
ahead. The systematic uncertainty in the calibrating SNe and their
Cepheid distances is still an important consideration, discussed in
the next section.

\section{Discussion}

\subsection{Comparison with other work}

\markcite{sun99}Suntzeff et al. (1999) used SN 1998bu and four other
local calibrators (they included SN 1937C in IC 4182 in their local
sample), combined with distant SNe Ia from the Cal\'an/Tololo sample
to derive a Hubble constant, $H_0 = 64.0 \pm 2.2 \; {\rm km \; s^{-1}
\; Mpc^{-1}}$ (this uncertainty ignores covariances in the Cepheid
distances). Their result, using a different method to convert light
curves to luminosities and extinction \markcite{phi99}(Phillips et
al. 1999), is quite consistent with ours. There are slight differences
in measurements of individual objects, but the overall agreement is
reassuring and indicates that SNe Ia are excellent distance indicators
whose intrinsic diversity can be understood and quantified.

The calibration of the peak absolute magnitude of SNe Ia has been
driven by the great efforts of the \emph{HST} program to measure
Cepheid distances to the supernova host galaxies
\markcite{san92}(Sandage et al. 1992). That group's latest published
determination of the Hubble constant and statistical uncertainty (also
ignoring Cepheid covariances) is $H_0 = 58 \pm 3 \; {\rm km \; s^{-1}
\; Mpc^{-1}}$ \markcite{sah97}(Saha et al. 1997). This is consistent
with our result -- agreement to $\sim$ 10\% is good given the long
history of measurements of the Hubble constant. Nonetheless, it is
instructive to pinpoint where the differences arise. The
\markcite{sah97}Saha et al. (1997) analysis uses a ``fiducial sample''
of 56 Hubble-flow SNe Ia with $B$ and $V$ peak magnitudes, typically
determined from photographic plates. Enforced upon the sample is a
velocity constraint, $3 < \log cz [{\rm km \; s^{-1}}] < 4.5$, and a
color constraint $-0.25 \leq B_{\rm max} - V_{\rm max} \leq 0.20$ mag
(after correction for Galactic extinction) to avoid peculiar SNe Ia
and those with large amounts of extinction. As local calibrators, they
use seven SNe Ia in six galaxies with five Cepheid distances (the
distance to SN 1989B was estimated by association to other Leo Group
galaxies with measured Cepheid distances; see their Table 6). Their
local calibrator sample is \emph{not} selected by the same criteria as
their fiducial sample, as no color constraint was applied to the local
calibrators. So SN 1895B was used with an estimated $B_{\rm max}$ but
no color information, and SN 1989B was used though it was too red,
with $B_{\rm max}-V_{\rm max} = 0.35 \pm 0.07$ mag
\markcite{wel94}(Wells et al. 1994). (SN 1998bu would also be too red
to meet the color requirement of their fiducial sample.) Nevertheless,
excluding SN 1895B and 1989B from their analysis would only have a
small effect and would increase their derived $H_0$ by about 1 ${\rm
km \; s^{-1} \; Mpc^{-1}}$.

\begin{figure*}
\plotone{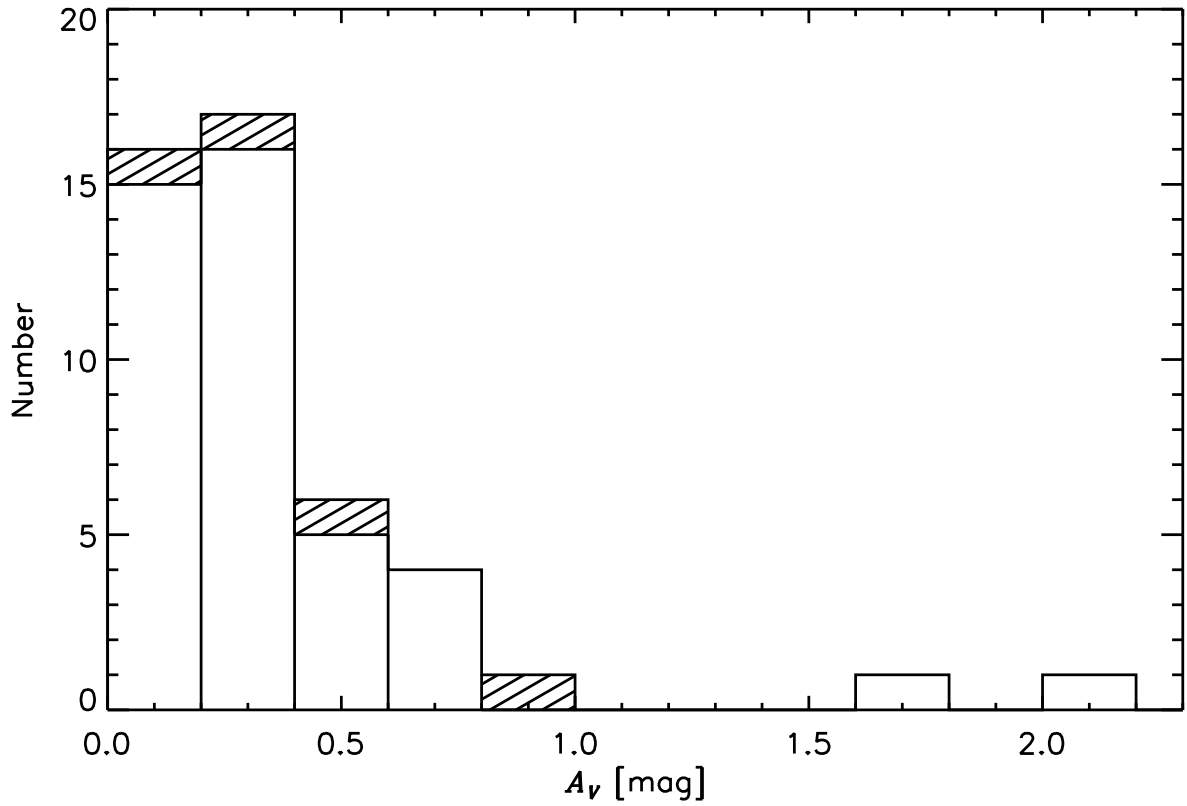}
\caption{Distribution of the total visual extinction, $A_V$, along the
lines-of-sight to the 42 Hubble-flow SNe Ia (\emph{clear}) and the 4
calibrators (\emph{shaded}). The extinction estimates are derived from
the MLCS analysis of the $BVRI$ light curves. \label{figav}}
\end{figure*}

Though we have not used SN 1895B and SN 1989B in our local sample for
reasons adduced earlier, there are additional differences between the
two analyses. \markcite{sah97}Saha et al. (1997) include no correction
for variation in intrinsic SN Ia luminosity based on light curve
shape. This is particularly significant for the calibrator sample
where 3 out of the 4 objects are slow decliners; including this
correction makes the estimated fiducial absolute peak magnitude
slightly fainter and explains about half our disagreement in
$H_0$. Most of the balance of the difference likely arises from the
treatment of extinction. While \markcite{sah97}Saha et al. (1997)
correct some of their local calibrator peak magnitudes for extinction
individually, they do not apply an extinction correction to their
fiducial sample SNe. If the mean color excess of their fiducial sample
were as little as $E(B-V) \simeq 0.03$ mag, correcting for extinction
would increase the mean $m_V$ by $\sim 0.1$ mag, and raise $a_V$ by
$\sim 0.02$, which is the other half of the difference in
$H_0$. \markcite{sah97}Saha et al. (1997) argue that selection effects
against the discovery of extinguished distant supernovae preclude a
significant amount of extinction in the fiducial sample. While this
\emph{could be} true, there is no demonstration that it \emph{is} true
for the sample they use. The selection effects in the several searches
that led to the SNe Ia of their fiducial sample are quite complicated
(see, e.g. \markcite{ham99}Hamuy \& Pinto 1999), and \emph{a priori}
statements about the possible extinction distribution of the distant
supernovae are not, by themselves, evidence. In particular, discovery
of SNe Ia with $A_V \simeq 0.1$ mag does not seem to be strongly
suppressed. In Figure \ref{figav} we show the extinction distribution
for our Hubble-flow and calibrating samples and it is clear that a
some of the supernovae found this way are, in fact, signifcantly
extinguished.

The MLCS analysis was designed to address these concerns; the
supernova peak magnitude, extinction and luminosity correction are
quantitatively estimated for each object, with a careful attention to
correlations and the final distance modulus uncertainty. This obviates
the need to make arbitrary sample cuts. We have taken particular care
to analyze the local calibrators and the Hubble-flow SNe Ia by the
same methods.  While these technical differences are significant, the
difference in the derived Hubble constant between our approach and
others is small. Uncertainty in the true Hubble constant still arises
principally from the small size of the calibrator sample and from
uncertainties in the lower rungs of the distance ladder.

\subsection{External uncertainties \label{secextunc}}

We examine three sources of systematic uncertainty in the distances of
the galaxies that host the calibrating SNe Ia: the effect of metallicity
on \emph{HST} Cepheid distance moduli, recalibration of the Cepheid
$PL$ relation, and the distance to the LMC, which are likely to be the
most important sources of systematic error.

\subsubsection{Metallicity dependence of the Cepheid scale}

The remarkably tight $PL$ relation of Cepheids in the LMC is quite a
boon to distance measurement, but also calls for understanding
variation in the $PL$ (and therefore derived distance moduli) with
environment. Metallicity, in particular, may play an important
role. Theoretical studies of Cepheid pulsation
\markcite{chi93}(e.g., Chiosi, Wood, \& Capitanio 1993) indicate
metallicity can have an effect on the brightness and colors of
Cepheids, though the size of this effect is uncertain.  If the
brightness and colors of Cepheids vary with metallicity, their
distances will be misestimated, both because of an incorrect estimate
of their intrinsic brightness as well as an incorrect estimate of the
extinction based on the color excess. Systematic errors in distances
would occur in Cepheid populations with significantly different mean
abundances than the LMC.

While it is important to understand the effects of metallicity on
Cepheid luminosities and colors in many wavebands
\markcite{fre90}\markcite{gou94}\markcite{sti95}(see, e.g., Freedman
\& Madore 1990; Gould 1994; Stift 1995), our particular concern is the
effect on the \emph{HST} Cepheid distances. Recently great efforts
have been made to determine empirically the relation between
metallicity and distance moduli measured in $V$ and $I$ with the
standard procedure of extinction correction. The
``metallicity'' dependence of extragalactic Cepheids is usually
parameterized in terms of [O/H], the logarithmic number abundance of
oxygen to hydrogen, relative to solar composition and measured via
\ion{H}{2} region spectra.  We adopt the notation
\begin{equation}
\gamma_{VI} = \frac{\Delta \mu}{\Delta [{\rm O}/{\rm H}]},
\end{equation}
which gives the change in the distance modulus per factor of ten in
metallicity, measured in mag/dex. Then the true distance modulus of a
galaxy is given by $\mu_{\rm true} = \mu_{VI} - \gamma_{VI} ([{\rm
O}/{\rm H}] - [{\rm O}/{\rm H}]_{\rm LMC})$, where $\mu_{VI}$ is the
measured extinction-corrected distance modulus. Recently,
\markcite{ken98}Kennicutt et al. (1998) used \emph{HST} observations
of three fields in M101 over which a large abundance gradient has been
observed from measurements of \ion{H}{2} regions. Their analysis
showed a slight correlation of distance modulus with metallicity, with
$\gamma_{VI} = -0.24 \pm 0.16$. \markcite{bea97}Beaulieu et al. (1997)
and \markcite{sas97}Sasselov et al. (1997) analyzed the effects of
metallicity on the $PL$ relation from observations of LMC and SMC
Cepheids from the EROS microlensing project. Their conclusion was for
a slightly stronger dependence\footnote{The value of $\gamma_{VI} =
-0.44$ quoted in their papers is for abundances measured in terms of
[Fe/H]; $\gamma_{VI} = -0.48$ is correct for [O/H] (D. Sasselov,
personal communication).}, $\gamma_{VI} = -0.48^{+0.1}_{-0.2}$. A
global analysis of many Galactic and extragalactic Cepheids by
\markcite{koc97}Kochanek (1997) yielded a similar metallicity
dependence; we consider this analysis in more detail in the next
section.

To assess the possible impact of metal abundance on the Hubble
Constant from SNe Ia, we have recalculated the distance moduli to the
calibrating galaxies with these empirically determined metallicity
corrections. We adopt $[{\rm O}/{\rm H}] - [{\rm O}/{\rm H}]_{\rm
LMC}$ measurements for NGC 3368 and NGC 5253 as compiled by
\markcite{ken98}Kennicutt et al. (1998) in their Table 4. The
abundances for NGC 4639 and NGC 4536 are from \markcite{koc97}Kochanek
(1997), who estimated these based on metallicity-magnitude and
metallicity-galaxy type relations; we have adopted an uncertainty of
$\pm 0.20$ on these estimates, though the results are not particularly
sensitive to this choice. The effect of the recalculated distance
moduli is shown in Table \ref{tabmetal}, where we have derived
$M^0_V$ estimates for three cases: $\gamma_{VI} \equiv 0$, i.e., no
metallicity dependence, $\gamma_{VI} = -0.24 \pm 0.16$ from
\markcite{ken98}Kennicutt et al. (1998), and $\gamma_{VI} =
-0.48^{+0.1}_{-0.2}$ from \markcite{beau97}Beaulieu et al. (1997) and
\markcite{sas97}Sasselov et al. (1997).

\begin{deluxetable}{llcccc}
\footnotesize
\tablecaption{Effects of Metallicity \label{tabmetal}}
\tablehead{
\colhead{} & \colhead{} & \colhead{} & \multicolumn{3}{c}{$M^0_V$} \\
\cline{4-6} 
\colhead{SN Ia} & \colhead{Galaxy} & \colhead{$[O/H] -
[O/H]_{\rm LMC}$} & \colhead{$\gamma_{VI}
\equiv 0$} & \colhead{$\gamma_{VI} = -0.24 \pm 0.16$} &
\colhead{$\gamma_{VI} = -0.48^{+0.1}_{-0.2}$} }
\startdata
1998bu	& NGC 3368 & $+0.70 \pm 0.20$ & $-19.44 \pm 0.22$
	& $-19.61 \pm 0.25$ & $-19.78^{+0.25}_{-0.27}$ \nl

1990N  	& NGC 4639 & $+0.10 \pm 0.20$ & $-19.46 \pm 0.33$
	& $-19.48 \pm 0.33$ & $-19.51^{+0.34}_{-0.34}$ \nl

1981B	& NGC 4536 & \phs$0.00 \pm 0.20$ & $-19.12 \pm 0.23$
	& $-19.12 \pm 0.23$ & $-19.12^{+0.24}_{-0.24}$ \nl

1972E	& NGC 5253 & $-0.35 \pm 0.15$ & $-19.42 \pm 0.29$
	& $-19.34 \pm 0.30$ & $-19.25^{+0.30}_{-0.30}$ \nl

\cline{1-6}

Mean	& \nodata & \nodata & $-19.34 \pm 0.17$ 
	& $-19.36 \pm 0.17$ & $-19.40 \pm 0.18$ \nl

$H_0$	& \nodata & \nodata & $64.4^{+5.6}_{-5.1}$
	& $63.8^{+5.6}_{-5.1}$ & $62.7^{+5.7}_{-5.2}$ \nl

\enddata
\end{deluxetable}

The results are interesting; the corrections for metallicity yield
little change in the inferred Hubble constant. Low $H_0$ values from
SNe Ia are not due to metallicity effects on the Cepheid distances to
this sample of galaxies. This is primarily due to our inclusion of SN
1998bu in NGC 3368, which is the only metal-rich calibrator, by a
factor of 0.7 dex compared to the LMC. If metallicity were the culprit
causing a low SNe Ia $H_0$, we would expect our inferred $M^0_V$ for
SN 1998bu to be very faint without a metallicity correction. However,
the distance modulus of \markcite{tan95}Tanvir et al. (1995) and our
observations do not require a faint value. More strikingly, comparing
SN 1998bu and SN 1972E we see that the estimates of $M^0_V$ are almost
exactly the same without a metallicity correction, even though the
metal abundance is higher in NGC 3368 than in NGC 5253 by more than a
factor 10.

As a result, including a metallicity correction increases the
dispersion in the estimates of $M^0_V$ from the four calibrating SNe
Ia, from $\sigma(M^0_V) = 0.16$ mag with no metallicity correction, to
$\sigma(M^0_V) = 0.21$ mag for the \markcite{ken98}Kennicutt et
al. (1998) value, to $\sigma(M^0_V) = 0.29$ mag for the
\markcite{sas97}Sasselov et al. (1997) value. The sample size of four
is probably too small to place much confidence in this result, but if
it were borne out by a larger sample, it would have interesting
implications. Assuming that the intrinsic dispersion in the
MLCS-corrected luminosity for the calibrating supernovae is the same
as that for the Hubble-flow SNe Ia ($\sigma \simeq 0.16$ mag), any
increase in the dispersion would arise from the Cepheid distance
moduli. A significant increase in dispersion with a metallicity
correction would imply either that the association between \ion{H}{2}
region metallicity and Cepheid metallicity is not straightforward,
that the metallicity correction was incorrect, or that some other
systematic uncertainty (perhaps in the properties of SNe in regions of
different metallicity) was colluding with the metallicity to counter
its effect in the uncorrected distance moduli. Increasing the sample
of SNe Ia in galaxies with Cepheid distances (particularly covering a
wide range of metallicity) would be a very desirable path to
understanding this important uncertainty in the distance scale. We
note that \markcite{nev98}Nevalainen \& Roos (1998) used this idea of
``statistical consistency'' between Cepheid-calibrated distance
indicators (including SNe Ia) to derive a metallicity dependence which
brought the estimates of $H_0$ into the best concordance, $\gamma_{VI}
= -0.31^{+0.15}_{-0.14}$. Because of the high metallicity of NGC 3368,
including SN 1998bu in their analysis would likely affect this result.

In principle the supernovae themselves can provide an estimate of the
metallicity correction, but minimizing the dispersion in our four
calibrating supernovae yields $\gamma_{VI} = +0.11 \pm 0.37$, which is
only a weak constraint given the very small sample size. In addition,
such a procedure ignores a possibly significant metallicity dependence
in the brightness of the supernovae themselves. Thus the current
supernova data do not provide strong evidence either for or against
the incorporation of a metallicity dependence in the Cepheid distance
moduli. We take the results at face value, and combined with the
current best estimates of $\gamma_{VI}$ we conclude that the
systematic $1\sigma$ error in $H_0$ from metallicity considerations
for this sample is small, $^{+0.0}_{-1.7} \; {\rm km \; s^{-1} \;
Mpc^{-1} }$.

\subsubsection{The Cepheid $PL$ relation}

We (and most authors) have adopted the $V$-band and $I$-band $PL$
relations derived from LMC Cepheids by \markcite{mad91}Madore \&
Freedman (1991). The relations are consistent with earlier estimates
(e.g., Sandage \& Tammann 1968; Feast \& Walker 1987), but are based
on a relatively small sample of objects compared to the number now
known in more distant galaxies. An error in the $PL$ slope or
zero-point could be an important source of systematic uncertainty. In
this section, by the zero-point we do not mean to include the
uncertainty in the distance to the LMC (still adopted as $\mu_{\rm
LMC} \equiv 18.50$, and discussed extensively in the next section),
but only uncertainty in the zero-point that arises from a small sample
of LMC Cepheids, even after assuming the LMC distance is perfectly
known.

The \markcite{mad91}Madore \& Freedman (1991) $PL$ relations have
statistical zero-point uncertainties of 0.05 and 0.03 mag in $V$ and
$I$ respectively. Unfortunately, residuals for the fit to the $PL$
relations are correlated, and the $PL$ slope is also slightly
correlated to the zero point, so that the real uncertainty in the
application of the $PL$ relations is not just a straightforward
quadrature combination of these uncertainties. \markcite{tan96}Tanvir
(1996) analyzed an augmented sample of LMC Cepheids with particular
attention to correlated residuals and found a slightly tighter
relation. However, his result indicated bias in the
\markcite{mad91}Madore \& Freedman (1991) calibration such that for
typical period distributions, \emph{HST} Cepheid distance moduli are
overestimated by $\sim 0.1$ mag, which would imply an increase in our
estimate of $M^0_V$ (to fainter intrinsic luminosity) by the same
amount and a $\sim 5\%$ increase in $H_0$ for any given LMC distance.

The global analysis of \markcite{koc97}Kochanek (1997) carefully and
consistently treats the Cepheid data to allow for distance estimates
including the systematic uncertainties we have discussed so
far. Again, the distance to the LMC is fixed at $\mu \equiv 18.5$. In
Table \ref{tabkoch} we present our estimates of $M^0_V$ given the
distance moduli derived by \markcite{koc97}Kochanek (1997) and
presented in his Table 3. We have added 0.05 mag to the NGC 3368
distance modului presented there to correct for the WFPC2
``long/short'' exposure effect, which was not included in his analysis
of the Cepheids in that galaxy (this is just a first approximation;
the correct method would be to repeat his analysis with the fainter
NGC 3368 Cepheid magnitudes). With that caveat, we have considered two
of his models, the first being Model 0 which derives distances in the
``standard'' method, with a global solution for the $PL$ relation and
correct treatment of correlated errors. As Table \ref{tabkoch} shows,
the $M^0_V$ estimates are generally fainter in this model, in line
with the suggestion of \markcite{tan96}Tanvir (1996), leading to a
modest increase in $H_0$. The second model we consider is Model 3-15,
which also derives a global solution, but further allows for effects
of metallicity and positive extinction. The result is again generally
to decrease the host galaxy distances, and lead to a fainter $M^0_V$
and higher $H_0$. However, we note that the estimates of $M^0_V$ are
not consistent with their uncertainties (which are derived from the
quadrature sum of the uncertainty in $m^0_V$ from MLCS and the
uncertainty in the Cepheid distance modulus). Furthermore, the
dispersion is much larger than the $\sim 0.16$ mag expected from the
SNe Ia alone; this implies that either the local calibrating SNe Ia
are very different from the Hubble-flow SNe Ia (a possibility we feel
is unlikely based on their spectra and light curves), or that
systematic errors remain in the Cepheid distances derived from this
particular model. Further analysis is required.

\begin{deluxetable}{llccc}
\tablecaption{Results Based on the Global Analysis of Kochanek (1997) 
\label{tabkoch}}
\tablehead{
\colhead{} & \colhead{} & \multicolumn{3}{c}{$M^0_V$} \\
\cline{3-5} 
\colhead{SN Ia} & \colhead{Galaxy} & \colhead{Published\tablenotemark{a}} &
\colhead{Model 0\tablenotemark{b}} & \colhead{Model 3-15\tablenotemark{b}} }
\startdata
1998bu	& NGC 3368 & $-19.44 \pm 0.22$
	& $-19.40 \pm 0.33$ & $-19.54 \pm 0.23$ \nl

1990N  	& NGC 4639 & $-19.46 \pm 0.33$
	& $-19.54 \pm 0.41$ & $-19.21 \pm 0.28$ \nl

1981B	& NGC 4536 & $-19.12 \pm 0.23$
	& $-18.99 \pm 0.28$ & $-18.63 \pm 0.22$ \nl

1972E	& NGC 5253 & $-19.42 \pm 0.29$
	& $-19.04 \pm 0.35$ & $-18.74 \pm 0.20$ \nl

\cline{1-5}

Mean	& \nodata & $-19.34 \pm 0.17$ 
	& $-19.20 \pm 0.22$ & $-19.03 \pm 0.27\tablenotemark{c}$ \nl

$H_0$	& \nodata & $64.4^{+5.6}_{-5.1}$
	& $68.7^{+7.6}_{-6.8}$ & $74.3^{+10.1}_{-8.9}$ \nl

\enddata
\tablenotetext{a}{Published refers to $M^0_V$ estimates based on
published distance moduli as described and modified in section 
\ref{sectcephcal}.}
\tablenotetext{b}{These two columns are from Kochanek (1997). See text
for details regarding the models.}
\tablenotetext{c}{Since the values in this column seem to be
inconsistent with their derived uncertainties, we have calculated an
unweighted mean.}
\end{deluxetable}

These results suggest a $1\sigma$ systematic uncertainty in the
calibrating SNe Ia host galaxy Cepheid distance moduli of
$^{+0.05}_{-0.10}$ mag due to the combined effects of metallicity,
extinction and the calibration of the $PL$ relation. (The resulting
uncertainty in $M^0_V$ is in the opposite sense, $^{+0.10}_{-0.05}$
mag.) Since definitive results regarding the exact magnitude of these
effects are lacking, we just include these effects in our uncertainty,
with $M^0_V = -19.34^{+0.20}_{-0.18}$. Our estimate of the Hubble
constant then becomes
\begin{equation}
H_0 = 64.4^{+6.6}_{-5.4} \; {\rm km \; s^{-1} \; Mpc^{-1} }.
\label{eqnh0sys}
\end{equation}

\subsubsection{Distance to the LMC\label{sectlmcdist}}

The final source of systematic error in the Hubble constant that we
consider is the distance to the LMC. All methods to measure
$H_0$ which are based on \emph{HST} Cepheid distances share this systematic
uncertainty, so that comparisons in the resulting $H_0$ values between
these methods should not include an error component from the LMC
distance (i.e., comparisons of \emph{HST} Cepheid-calibrated $H_0$
measurements should be to the analog of equation \ref{eqnh0sys}; if
two such measurements disagree, they will disagree regardless of the
LMC distance).  Formally, our best estimate of the Hubble constant,
including the systematic uncertainties discussed above, is given by
\begin{equation}
\log H_0 = 1.809^{+0.042}_{-0.038} - 0.2\left(\mu_{\rm LMC} -
18.50\right).
\label{eqnh0plmc}
\end{equation}
Of course, if we wish to compare our value of $H_0$ with those derived
from techniques which are independent of the LMC distance (e.g., SNe
II expanding photospheres, gravitational lens time delays,
Sunyaev-Zel'dovich effect, etc.), we have to provide a best estimate
for $\mu_{LMC}$ and perhaps more importantly, its
uncertainty. Measurement of quantities derived from the ``true''
Hubble constant, such as the dynamical age of the Universe, also
requires such an estimate.

\begin{figure*}
\plotone{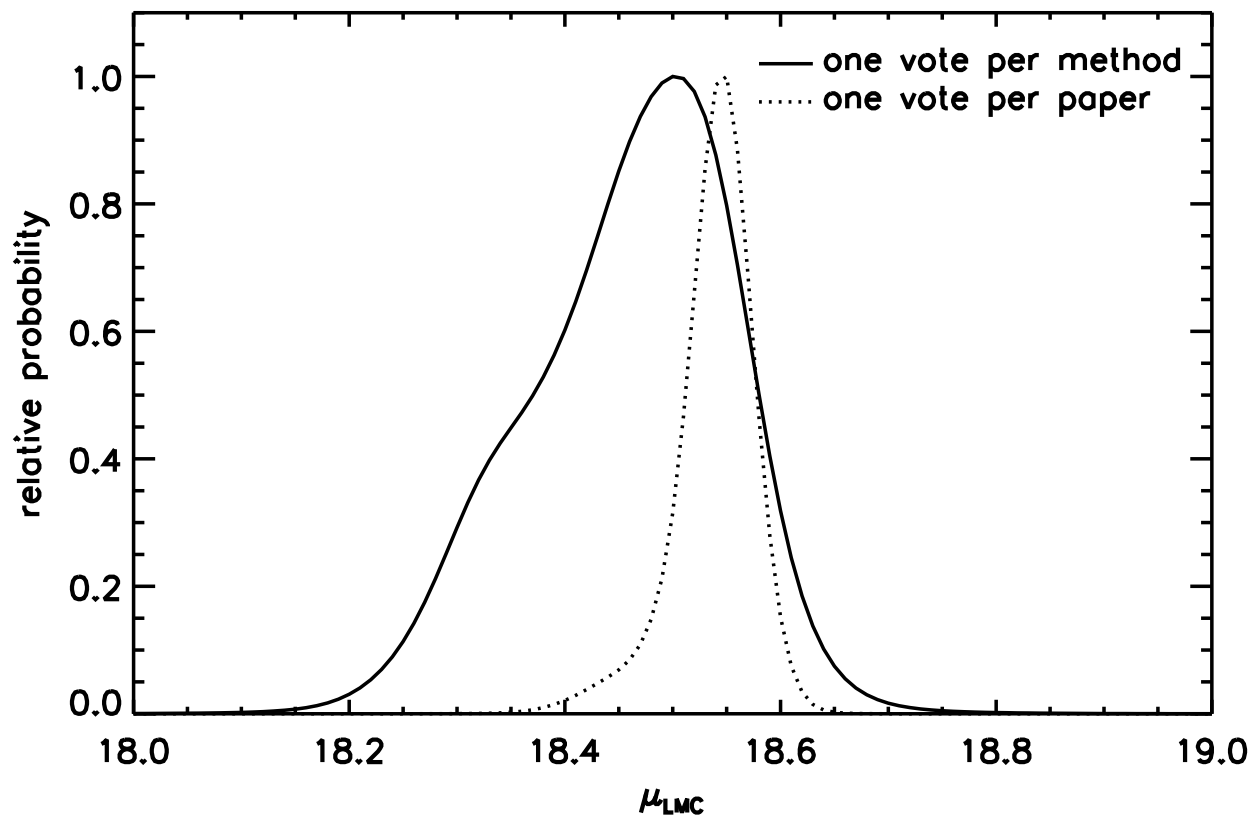}
\caption{Probability density functions for the mean LMC distance
modulus based on a Bayesian analysis of 19 recent measurements. The
dotted curve shows the pdf if each measurement is treated
independently, while the solid curve shows the PDF when each distance
measuring \emph{technique} is given equal weight. For clarity, both
distributions have been rescaled to peak at unity. \label{figlmcdist}}
\end{figure*}

However, measurements of the LMC distance modulus are notoriously
inconsistent, ranging for the most part from $\sim18.2$ to $\sim
18.7$. The value of $\mu_{\rm LMC} = 18.5$ that we have adopted has
recently faced a strong challenge from a ``short'' LMC distance based
on Hipparcos-calibrated red clump stars
\markcite{sta98}\markcite{uda98}(Stanek, Zaritsky, \& Harris 1998;
Udalski 1998) and the study of detached eclipsing binaries such as
HV2274 \markcite{gui98}(Guinan et al. 1998), which give $\mu_{\rm LMC}
= 18.18 \pm 0.06$ and $18.30 \pm 0.07$, respectively. However, recent
applications of other methods, including Cepheids, RR Lyrae stars,
Mira variables, and the SN 1987A ring still yield a wide range of
distance moduli, many with inconsistent error bars.  As an exercise,
we compiled a representative (though not exhaustive) sample of 19 LMC
distance moduli published in the last two years
\markcite{alc97}\markcite{dib97}\markcite{fea97}\markcite{grat97}\markcite{pan97}\markcite{van97}\markcite{whi97}\markcite{ber98}\markcite{col98}\markcite{fer98}\markcite{gie98}\markcite{gou98}\markcite{gui98}\markcite{lur98a}\markcite{lur98b}\markcite{madf98}\markcite{oud98}\markcite{rei98}\markcite{uda98}
(Alcock et al. 1997; Di Benedetto 1997; Feast \& Catchpole 1997;
Gratton et al. 1997; Panagia et al. 1997; van Leeuwen et al. 1997;
Whitelock, van Leeuwen, \& Feast 1997; Bergeat, Knapik, \& Rutily
1998; Cole 1998; Fernley et al. 1998; Gieren, Fouqu\'e, \& G\'omez
1998; Gould \& Uza 1998; Guinan et al. 1998; Luri et al. 1998a; Luri
et al. 1998b; Madore \& Freedman 1998; Oudmaijer, Groenewegen, \&
Schrijver 1998; Reid 1998; Udalski 1998). Naively assuming each
measurement to be independent, a Bayesian analysis of these distance
estimates in the spirit of \markcite{pre97}Press (1997) yields a
narrow probability density function (PDF) for the mean, with $\mu_{\rm
LMC} = 18.55^{+0.02}_{-0.04}$, as shown in Figure \ref{figlmcdist}. If
we modify the analysis so that each distance method only gets one
``vote'' to reduce correlated errors, the PDF becomes quite asymmetric
with $\mu_{\rm LMC} = 18.50^{+0.05}_{-0.15}$, where the stated value
is the peak of the PDF and the upper and lower uncertainties are
derived from the points at which the cumulative probability is 0.841
and 0.159. This analysis does not provide a reason for the discrepant
values, and is subject to additional correlated errors that may still
be lurking. Nevertheless, given the incompatible data, the method
provides a reasonable and statistically defensible way to estimate
$\mu_{\rm LMC}$, and its uncertainty.

As the current best estimate of the LMC distance modulus, we adopt
$\mu_{\rm LMC} = 18.50^{+0.10}_{-0.15}$ mag.  Others will undoubtedly
have differing estimates, and can use equation \ref{eqnh0plmc} to
determine the resulting Hubble constant. With our choice, we have as
the estimate of the fiducial absolute magnitude, including all
identified systematic uncertainties, $M^0_V =
-19.34^{+0.25}_{-0.21}$. Thus, our final estimate of the Hubble
constant incorporating this total systematic uncertainty is
\begin{equation}
H_0 = 64.4^{+8.1}_{-6.2} \; {\rm km \; s^{-1} \; Mpc^{-1}}.
\end{equation}

\subsection{Implications}

One direct implication of our derived Hubble constant is an estimate
of the dynamical age of the universe, $t_0$, assuming a
Friedmann-Robertson-Walker cosmology. In an $\Omega_M =
1$, Einstein-de Sitter universe, $H_0 t_0 =
2/3$. However, evidence from SNe Ia at high redshifts
provides a strong observational constraint on this product, with
\markcite{rie98a}Riess et al. (1998a) deriving $H_0 t_0 = 0.93 \pm
0.06$ and \markcite{per99}Perlmutter et al. (1999) obtaining $H_0
t_0 = 0.93 \pm 0.05$. Assuming that systematic errors in these
estimates are small (discussed extensively in both papers), we have
\begin{equation}
t_0 = 14.1 \pm 1.6 \; {\rm Gyr}.
\end{equation}
This estimate is very nearly the same as that presented by Riess et
al. \markcite{rie98a}(1998), and is in good accord with measurements
of the ages of the oldest objects in the Universe \markcite{cha98}(see
Chaboyer 1998 for a review). The solution to the cosmological ``age
crisis'' is not to be found in alternative estimates of $H_0$, but
rather in discarding the $\Omega_M = 1$ cosmology (which would require
$t_0 = 10.1 \pm 1.1$ Gyr).

The supernovae can also tell us about the structure of groups and
clusters. For instance, \markcite{gra97}Graham et al. (1997) obtained
a Cepheid distance to NGC 3351 which is also a member of the Leo I
group along with NGC 3368. However, the distance modulus they derive,
$\mu_{\rm NGC 3351} = 30.01 \pm 0.19$, is almost 0.4 mag closer than
the \markcite{tan95} Tanvir et al. (1995) distance for NGC 3368,
$\mu_{\rm NGC 3368} = 30.37 \pm 0.12$. This corresponds to a 2 Mpc
difference in the line of sight distance to these galaxies, even
though their projected separation is 41 arcmin, or about 120 kpc at
the inferred distance. \markcite{gra97}Graham et al. (1997) suggest
one alternative explanation is that the $I$-band NGC 3368 Cepheid
photometry of \markcite{tan95} Tanvir et al. (1995) is systematically
faint, which would lead to an underestimated extinction and an
overestimated distance. Metallicity corrections to the Cepheid
distances do not change this conclusion, as both NGC 3368 and NGC 3351
are similarly metal rich \markcite{ken98}(Kennicutt et al. 1998).  If
we adopt the mean $M^0_V$ from the three other calibrators, SN 1998bu
can in principle be used to test the Tanvir et al. (1995) distance.
However, this procedure yields $\mu_{\rm NGC 3368} = 30.26 \pm 0.27$,
and the uncertainty is too large to provide a definitive test.

This discussion does point out an important attribute of our
analysis. We have assumed that the internal errors in the Cepheid
distances (i.e., excluding those factors we discuss in Section
\ref{secextunc}) are accurately estimated. While we have tried to
create a uniform set of distance moduli from the published values, a
better procedure would be to reanalyze all the Cepheid data from the
different groups in a completely consistent manner, beginning with the
raw \emph{HST} images. The Key Project team is doing this
\markcite{gib99}(Gibson et al. 1999); if needed, revised distance
moduli for the SN calibrators can then be used in Table \ref{tabcalsn}
to determine a new estimate for $M^0_V$ and $H_0$.

More information about the structure of the Leo I group can be
determined by relying on the supernovae themselves. For instance, the
galaxy NGC 3389, host of the SN Ia 1967C, is sometimes considered a
member of the Leo I group, though various group-finding algorithms
disagree about its inclusion \markcite{schn89}(Schneider 1989). We
have used MLCS to analyze the photoelectric light curve of SN 1967C
\markcite{dev68}(de Vaucouleurs 1968), and conclude that NGC 3389 is
at $\mu \simeq 32.6$, placing it significantly farther away than the
other Leo I group galaxies.

We can also compare the distance of NGC 3368 with that of NGC 3627, in
the Leo triplet, located on the sky roughly 8 degrees from the Leo I
group. NGC 3627 was the host of the SN Ia 1989B, which had a
light curve similar to that of SN 1998bu. From our MLCS analysis of
the SN 1989B we derive $\langle V_{\rm Bmax} \rangle = 12.01$, $A_V =
0.99$ mag, and $\Delta = 0.19$ mag, yielding $m^0_V = 10.83 \pm
0.19$. Combined with our result for SN 1998bu, $m^0_V = 10.93 \pm
0.18$, we see that NGC 3627 is most likely $\sim 0.1$ mag closer than
NGC 3368, though there is a bit of uncertainty (arising from the
correlated uncertainties in the MLCS analysis of each
supernova). Since the supernovae only tell us the distances of their
hosts, without further information we cannot determine the
relationship between the Leo I Group and the Leo triplet as a whole.
\emph{HST} data has been taken to measure a Cepheid distance to NGC
3627 directly; with our analysis of SN 1989B and taking $M^0_V =
-19.34 \pm 0.17$ derived from all four calibrators, we predict the
distance modulus for NGC 3627, $\mu = 30.17 \pm 0.25$. In a very
recent preprint, \markcite{sah99}Saha et al. (1999) report their
Cepheid distance to NGC 3627, with the result $\mu = 30.22 \pm 0.12$,
in excellent agreement with our expectation. With this new Cepheid
distance, we will be able to use SN 1989B as a calibrator in the
future; including it in the analysis presented here would decrease our
estimate $H_0$ by merely 0.3 ${\rm km \; s^{-1} \; Mpc^{-1}}$.

This discussion strengthens our opinion that the sample of calibrating
SNe Ia should be restricted to those for which Cepheid distances have
been directly determined, rather than relying on indirect association
with other members of a group or cluster. \markcite{sun99}Suntzeff et
al. (1999) make this point with two SNe Ia in the Fornax cluster, SN
1980N in NGC 1316 and SN 1992A in NGC 1380. Adopting the Cepheid
distance to NGC 1365 \markcite{sil99}\markcite{mad99}(Silbermann et
al. 1999; Madore et al. 1999) as the distance to the cluster and these
two galaxies in particular leads to a correlated systematic effect on
the SN Ia calibration. The analysis of Suntzeff et al. (1999)
indicates that NGC 1365 is probably foreground to both NGC 1316 and
NGC 1380 by $\sim 0.3$ mag. We have analyzed the SN Ia in both of
these galaxies with MLCS and concur that NGC 1365 is closer than NGC
1316 by $\sim 0.3$ mag. Our result for NGC 1380 is slightly different,
with the MLCS analysis suggesting NGC 1365 is closer than NGC 1380 by
only $\sim 0.05$ mag. Nevertheless, this is exactly the sort of
systematic error we wish to avoid, since the number of calibrators is
small and systematic errors in any of them can significantly affect
the mean.

\section{Conclusion}

We have presented extensive photometric and spectroscopic observations
of SN 1998bu as well as an MLCS analysis to determine the intrinsic
luminosity (relative to the fiducial) of the supernova and the
extinction along the line of sight. Using the Cepheid distance to NGC
3368 and three other SN Ia host galaxies, we have calibrated the
absolute magnitude of our fiducial SN Ia, and applied this calibration
to a set of 42 distant SNe Ia to derive a Hubble constant, $H_0 =
64^{+8}_{-6} \; {\rm km \; s^{-1} \; Mpc^{-1}}$, including systematic
uncertainties such as in the distance to the LMC. The statistical
uncertainty in our estimate ($\sim 0.17$ mag) arises from the fact
that we have only 4 local calibrators; this uncertainty will be reduced
by more SNe Ia occurring in galaxies with \emph{HST} Cepheid
distances, or more controllably by measuring more Cepheid distances to
the host galaxies of well-observed SNe Ia (SN 1998aq in NGC 3982 is an
excellent target, for example). Reducing the systematic uncertainty in
the Hubble constant ($\sim 0.15$ mag) will be more difficult as it
will entail a better understanding of the Cepheid distance scale and
most importantly, a definitive distance modulus for the
LMC. Nevertheless, we are optimistic that these reductions are
possible, and eagerly await the day in the not-too-distant future when
the Hubble constant is known to better than 10\%.

\acknowledgements

We gratefully acknowledge the efforts of the staff maintaining the
many telescopes we have utilized. We also thank those engaged in
searching the skies for new supernovae, and particularly the amateur
astronomers who dedicate themselves to the task. Thanks also to Dan
Green and Brian Marsden of the IAU Central Bureau for Astronomical
Telegrams, who play a pivotal role in disseminating information about
new supernova discoveries and enable quick follow-up. We are grateful
to Di Harmer and Richard Green for accomodating our WIYN Target of
Opportunity observations, as well as Bob Joseph for allowing us IRTF
observations during engineering time. We thank F. Patat for providing
absorption-line velocities for several SNe Ia in electronic format,
and M. Hanson for advice on reducing infrared spectra. We are grateful
to Chris Kochanek, Dimitar Sasselov and Alyssa Goodman for helpful
discussions, and to the referee, Nick Suntzeff, for suggesting
many improvements to the manuscript. P.J.B. thanks the W. M. Keck
Foundation for its support of astronomy at Wellesley College through
the Keck Northeast Astronomy Consortium and the Wellesley College
Brachman Hoffman Research Grant. E.K.G. gratefully acknowledges
support by Dennis Zaritsky through NASA LTSA grant NAG-5-3501 and by
NASA through grant HF-01108.01-98A from the Space Telescope Science
Institute, which is operated by the Association of Universities for
Research in Astronomy, Inc., under NASA contract NAS5-26555. This work
was also supported by NSF grants AST-9528899 (R.P.K.), AST-9417213
(A.V.F.), and AST-9417359 (P.J.B.), as well as through an NSF Graduate
Research Fellowship (S.J.).

\end{document}